\documentclass[twocolumn,prc,aps,showpacs,superscriptaddress,amsmath,amssymb,floatfix,nofootinbib,superscriptaddress]{revtex4}

\usepackage{amsmath}
\usepackage{bm}
\usepackage{dcolumn}
\usepackage{psfig}
\usepackage{subfigure}
\usepackage{amssymb}
\usepackage{xspace}
\usepackage{graphicx}
\usepackage{epsfig}
\usepackage{draftcopy}

\newcommand{\etal}{{\it et al.\/ }}

\renewcommand{\deg}{\ensuremath{^\circ}}

\def\be{\begin{equation}}
\def\ee{\end{equation}}
\def\bea{\begin{eqnarray}}
\def\eea{\end{eqnarray}}
\def\ie{i.e.}

\def\eg{e.g.}

\newcommand{\sigl}{\mbox{\ensuremath{\sigma_\mathrm{L}}}}
\newcommand{\sigt}{\mbox{\ensuremath{\sigma_\mathrm{T}}}}
\newcommand{\siglt}{\mbox{\ensuremath{\sigma_\mathrm{LT}}}}
\newcommand{\sigtt}{\mbox{\ensuremath{\sigma_\mathrm{TT}}}}
\newcommand{\sigL}{\mbox{\ensuremath{\sigma_\mathrm{L}}}}
\newcommand{\sigT}{\mbox{\ensuremath{\sigma_\mathrm{T}}}}
\newcommand{\sigLT}{\mbox{\ensuremath{\sigma_\mathrm{LT}}}}
\newcommand{\sigTT}{\mbox{\ensuremath{\sigma_\mathrm{TT}}}}
\newcommand{\eps}{\mbox{\ensuremath{\epsilon}}}
\newcommand{\qsq}{\mbox{\ensuremath{Q^2}}}
\newcommand{\fpi}{\mbox{\ensuremath{F_\pi}}}

\newcommand{\Lpi}{\mbox{\ensuremath{\Lambda^2_\pi}}}
\newcommand{\Lrho}{\mbox{\ensuremath{\Lambda^2_\rho}}}
\newcommand{\gevsq}{\mbox{\ensuremath{\text{GeV}\text{}^2}}}
\def\fpiq2{$F_{\pi}(Q^2)$}

\def\heepi{${^1}$H($e,e'\pi^+$)$n$}
\def\heep{${^1}$H($e,e^\prime p$)}

\newcommand{\phd}{Ph.D. thesis}

\newcommand{\privcom}{private communication}

\begin{document}

\title{Charged pion form factor between $Q^2$=0.60 and 2.45 \gevsq. I. Measurements of the cross section for the \heepi\ reaction}

\author{H.P. Blok}
\affiliation{Dept. of Physics, VU university, NL-1081 HV Amsterdam, The Netherlands}
\affiliation{NIKHEF, Postbus 41882, NL-1009 DB Amsterdam, The Netherlands}
\author{T. Horn}
\affiliation{University of Maryland, College Park, Maryland 20742}
\affiliation{Physics Division, TJNAF, Newport News, Virginia 23606}
\author{G.M. Huber}
\affiliation{University of Regina, Regina, Saskatchewan S4S 0A2, Canada}
\author{E.J. Beise}
\affiliation{University of Maryland, College Park, Maryland 20742}
\author{D. Gaskell}
\affiliation{Physics Division, TJNAF, Newport News, Virginia 23606}
\author{D.J. Mack}
\affiliation{Physics Division, TJNAF, Newport News, Virginia 23606}
\author{V. Tadevosyan}
\affiliation{Yerevan Physics Institute, 375036 Yerevan, Armenia}
\author{J. Volmer}
\affiliation{Faculteit Natuur- en Sterrenkunde, Vrije Universiteit, NL-1081 HV 
        Amsterdam, The Netherlands}
\affiliation{DESY, Hamburg, Germany}
\author{D. Abbott}
\affiliation{Physics Division, TJNAF, Newport News, Virginia 23606}
\author{K. Aniol}
\affiliation{California State University Los Angeles, Los Angeles, California
  90032}
\author{H. Anklin}
\affiliation{Florida International University, Miami, Florida 33119}
\affiliation{Physics Division, TJNAF, Newport News, Virginia 23606}
\author{C. Armstrong}
\affiliation{College of William and Mary, Williamsburg, Virginia 23187}
\author{J. Arrington}
\affiliation{Argonne National Laboratory, Argonne, Illinois 60439}
\author{K. Assamagan}
\affiliation{Hampton University, Hampton, Virginia 23668}
\author{S. Avery}
\affiliation{Hampton University, Hampton, Virginia 23668}
\author{O.K. Baker}
\affiliation{Hampton University, Hampton, Virginia 23668}
\affiliation{Physics Division, TJNAF, Newport News, Virginia 23606}
\author{B. Barrett}
\affiliation{Saint Mary's University, Halifax, Nova Scotia, Canada}
\author{C. Bochna}
\affiliation{University of Illinois, Champaign, Illinois 61801}
\author{W. Boeglin}
\affiliation{Florida International University, Miami, Florida 33119}
\author{E.J. Brash}
\affiliation{University of Regina, Regina, Saskatchewan S4S 0A2, Canada}
\author{H. Breuer}
\affiliation{University of Maryland, College Park, Maryland 20742}
\author{C.C. Chang}
\affiliation{University of Maryland, College Park, Maryland 20742}
\author{N. Chant}
\affiliation{University of Maryland, College Park, Maryland 20742}
\author{M.E. Christy}
\affiliation{Hampton University, Hampton, Virginia 23668}
\author{J. Dunne}
\affiliation{Physics Division, TJNAF, Newport News, Virginia 23606}
\author{T. Eden}
\affiliation{Physics Division, TJNAF, Newport News, Virginia 23606}
\affiliation{Norfolk State University, Norfolk, Virginia}
\author{R. Ent}
\affiliation{Physics Division, TJNAF, Newport News, Virginia 23606}
\author{H. Fenker}
\affiliation{Physics Division, TJNAF, Newport News, Virginia 23606}
\author{E. Gibson}
\affiliation{California State University, Sacramento, California 95819}
\author{R. Gilman}
\affiliation{Rutgers University, Piscataway, New Jersey 08855}
\affiliation{Physics Division, TJNAF, Newport News, Virginia 23606}
\author{K. Gustafsson}
\affiliation{University of Maryland, College Park, Maryland 20742}
\author{W. Hinton}
\affiliation{Hampton University, Hampton, Virginia 23668}
\author{R.J. Holt}
\affiliation{Argonne National Laboratory, Argonne, Illinois 60439}
\author{H. Jackson}
\affiliation{Argonne National Laboratory, Argonne, Illinois 60439}
\author{S. Jin}
\affiliation{Kyungpook National University, Taegu, Korea}
\author{M.K. Jones}
\affiliation{College of William and Mary, Williamsburg, Virginia 23187}
\author{C.E. Keppel}
\affiliation{Hampton University, Hampton, Virginia 23668}
\affiliation{Physics Division, TJNAF, Newport News, Virginia 23606}
\author{P.H. Kim}
\affiliation{Kyungpook National University, Taegu, Korea}
\author{W. Kim}
\affiliation{Kyungpook National University, Taegu, Korea}
\author{P.M. King}
\affiliation{University of Maryland, College Park, Maryland 20742}
\author{A. Klein}
\affiliation{Old Dominion University, Norfolk, Virginia 23529}
\author{D. Koltenuk}
\affiliation{University of Pennsylvania, Philadelphia, Pennsylvania 19104}
\author{V. Kovaltchouk}
\affiliation{University of Regina, Regina, Saskatchewan S4S 0A2, Canada}
\author{M. Liang}
\affiliation{Physics Division, TJNAF, Newport News, Virginia 23606}
\author{J. Liu}
\affiliation{University of Maryland, College Park, Maryland 20742}
\author{G.J. Lolos}
\affiliation{University of Regina, Regina, Saskatchewan S4S 0A2, Canada}
\author{A. Lung}
\affiliation{Physics Division, TJNAF, Newport News, Virginia 23606}
\author{D.J. Margaziotis}
\affiliation{California State University Los Angeles, Los Angeles, California
  90032}
\author{P. Markowitz}
\affiliation{Florida International University, Miami, Florida 33119}
\author{A. Matsumura}
\affiliation{Tohoku University, Sendai, Japan}
\author{D. McKee}
\affiliation{New Mexico State University, Las Cruces, New Mexico 88003-8001}
\author{D. Meekins}
\affiliation{Physics Division, TJNAF, Newport News, Virginia 23606}
\author{J. Mitchell}
\affiliation{Physics Division, TJNAF, Newport News, Virginia 23606}
\author{T. Miyoshi}
\affiliation{Tohoku University, Sendai, Japan}
\author{H. Mkrtchyan}
\affiliation{Yerevan Physics Institute, 375036 Yerevan, Armenia}
\author{B. Mueller}
\affiliation{Argonne National Laboratory, Argonne, Illinois 60439}
\author{G. Niculescu}
\affiliation{James Madison University, Harrisonburg, Virginia 22807}
\author{I. Niculescu}
\affiliation{James Madison University, Harrisonburg, Virginia 22807}
\author{Y. Okayasu}
\affiliation{Tohoku University, Sendai, Japan}
\author{L. Pentchev}
\affiliation{College of William and Mary, Williamsburg, Virginia 23187}
\author{C. Perdrisat}
\affiliation{College of William and Mary, Williamsburg, Virginia 23187}
\author{D. Pitz}
\affiliation{DAPNIA/SPhN, CEA/Saclay, F-91191 Gif-sur-Yvette, France}
\author{D. Potterveld}
\affiliation{Argonne National Laboratory, Argonne, Illinois 60439}
\author{V. Punjabi}
\affiliation{Norfolk State University, Norfolk, Virginia}
\author{L.M. Qin}
\affiliation{Old Dominion University, Norfolk, Virginia 23529}
\author{P. Reimer}
\affiliation{Argonne National Laboratory, Argonne, Illinois 60439}
\author{J. Reinhold}
\affiliation{Florida International University, Miami, Florida 33119}
\author{J. Roche}
\affiliation{Physics Division, TJNAF, Newport News, Virginia 23606}
\author{P.G. Roos}
\affiliation{University of Maryland, College Park, Maryland 20742}
\author{A. Sarty}
\affiliation{Saint Mary's University, Halifax, Nova Scotia, Canada}
\author{I.K. Shin}
\affiliation{Kyungpook National University, Taegu, Korea}
\author{G.R. Smith}
\affiliation{Physics Division, TJNAF, Newport News, Virginia 23606}
\author{S. Stepanyan}
\affiliation{Yerevan Physics Institute, 375036 Yerevan, Armenia}
\author{L.G. Tang}
\affiliation{Hampton University, Hampton, Virginia 23668}
\affiliation{Physics Division, TJNAF, Newport News, Virginia 23606}
\author{V. Tvaskis}
\affiliation{Faculteit Natuur- en Sterrenkunde, Vrije Universiteit, NL-1081 HV 
        Amsterdam, The Netherlands}
\author{R.L.J. van der Meer}
\affiliation{University of Regina, Regina, Saskatchewan S4S 0A2, Canada}
\author{K. Vansyoc}
\affiliation{Old Dominion University, Norfolk, Virginia 23529}
\author{D. Van Westrum}
\affiliation{University of Colorado, Boulder, Colorado 76543}
\author{S. Vidakovic}
\affiliation{University of Regina, Regina, Saskatchewan S4S 0A2, Canada}
\author{W. Vulcan}
\affiliation{Physics Division, TJNAF, Newport News, Virginia 23606}
\author{G. Warren}
\affiliation{Physics Division, TJNAF, Newport News, Virginia 23606}
\author{S.A. Wood}
\affiliation{Physics Division, TJNAF, Newport News, Virginia 23606}
\author{C. Xu}
\affiliation{University of Regina, Regina, Saskatchewan S4S 0A2, Canada}
\author{C. Yan}
\affiliation{Physics Division, TJNAF, Newport News, Virginia 23606}
\author{W.-X. Zhao}
\affiliation{M.I.T.--Laboratory for Nuclear Sciences and Department of Physics, 
      Cambridge, Massachusetts 02139}
\author{X. Zheng}
\affiliation{Argonne National Laboratory, Argonne, Illinois 60439}
\author{B. Zihlmann}
\affiliation{University of Virginia, Charlottesville, Virginia 22901}
\affiliation{Physics Division, TJNAF, Newport News, Virginia 23606}
\collaboration{The Jefferson Lab \fpi\ Collaboration}
\noaffiliation

\date{\today}
 
\begin{abstract}
Cross sections for the reaction \heepi\ were measured in Hall C at 
Thomas Jefferson National
Accelerator Facility (JLab) using the CEBAF high-intensity, continous
electron beam in order to determine the charged pion form factor.
Data were taken for central four-momentum transfers ranging from $Q^2$=0.60 to
2.45 GeV$^2$ at an invariant mass of the virtual photon-nucleon system of $W$=1.95 and 2.22 GeV.
The measured cross sections were separated into the four structure functions
\sigl, \sigt, \siglt, and \sigtt.
The various parts of the experimental setup and the analysis steps are described in detail,
including the calibrations and systematic studies, which were needed to obtain high precision results.
The different types of systematic uncertainties are also discussed.
The results for the separated cross sections as a function of the Mandelstam variable $t$
at the different values of \qsq\ are presented. Some global features of the data are
discussed, and the data are compared with the results of some model calculations
for the reaction \heepi.
\end{abstract}


\maketitle

\newpage

\newpage


\section{Introduction}
\label{sec:intro}

A fundamental challenge in hadronic physics is trying to understand the
structure of mesons and baryons in terms of their quark-gluon
constituents, as given by the underlying theory of the strong interaction. This
theory is known as Quantum Chromodynamics (QCD).  Form factors of
hadrons play an important role in this description, because they provide
information about the internal structure of the hadron.

One of the simplest hadronic systems available for study is the pion, whose
valence structure is a bound state of a quark and an antiquark. Its
electromagnetic structure is parameterized by a single form factor, \fpiq2,
which depends on $Q^2=-q^2$, where $q^2$ is the four-momentum squared of the virtual 
photon. $F_{\pi}$ is well determined up to values of $Q^2$ of
0.28 GeV$^2$ by elastic $\pi-e$ scattering~\cite{ady77,dal82,ame86},
from which the charge radius has been extracted.
Determining \fpiq2\ at larger values of $Q^2$ requires the use of pion
electroproduction from a nucleon target. The longitudinal part of the cross
section for pion electroproduction, \sigL, contains the pion exchange
($t$-pole) process, in which the virtual photon couples to a virtual pion
inside the nucleon. This process is expected to dominate at small values of the
Mandelstam variable $-t$, thus allowing for the determination of \fpi.

Pion electroproduction data have previously been obtained for values of \qsq\
of 0.18 to 9.8 \gevsq\ at the Cambridge Electron Accelerator (CEA) and at
Cornell~\cite{beb76, beb78}, and at the Deutsches Elektronen-Synchrotron
(DESY)~\cite{bra77, ack78}. Most of the high $Q^2$ data have come from
experiments at Cornell. In these experiments, $F_{\pi}$ was extracted from the
longitudinal cross sections, which were isolated by subtracting a model of the
transverse contribution from the unseparated cross sections.
Pion electroproduction data were also obtained at DESY~\cite{ack78,bra76,bra77}
for values of $Q^2$ of 0.35 and 0.7 GeV$^2$, and longitudinal (L) and transverse
(T) cross sections were extracted using the Rosenbluth L/T separation method.

With the availability of the high-intensity, continuous electron beams and
well-understood magnetic spectrometers at the Thomas Jefferson National
Accelerator Facility (JLab) it became possible to determine L/T separated cross
sections with high precision, and thus to study the pion form factor in the
regime of \qsq=0.5-3.0 \gevsq.
In 1997, high-precision pion electroproduction data for values of $Q^2$ between
0.60 and 1.60 \gevsq\ were acquired at JLab at a value of the
invariant mass of the photon-nucleon system of $W$=1.95 GeV. 
The main results were published in  \cite{vol01}, with an updated analysis
of these data published in  \cite{tad06}. 
In 2003, the range $Q^2$=1.60-2.45 \gevsq\ was studied at a
value $W$=2.22 GeV~\cite{horn06}.  At each value of $Q^2$, cross sections were
obtained at two different values of the virtual photon polarization,
$\epsilon$, allowing for the separation of the longitudinal and transverse
components of the cross section.

The purpose of this work is to describe the experiment and analysis
in detail 
and to present and discuss additional results.
The discussion has been split into two parts. This paper
describes the experiment and analysis, presents the measured cross sections,
including the separation into the structure functions, along with a detailed
discussion of the systematic uncertainties, and compares them with previous 
L/T separated data and with theoretical calculations for the cross section. 
The paper immediately following \cite{hub08} discusses the
determination of \fpi\ and presents the resulting \fpi\ values, including all
uncertainties. These values are then compared to various theoretical predictions. 
This division was chosen to separate the determination
of the cross section, with its various experimental issues,
from the extraction of \fpi\ from the measured cross sections, 
which is model dependent.
If more advanced or other models will become available, new values for \fpi\
may be extracted from the same cross sections.

This paper is organized as follows: In the next section the basic formalism of
the \heepi\ reaction is presented. In section~\ref{sec:experiment} the
experiment performed at JLab is described, including the experimental set-up
and the calibrations of the spectrometers. The data analysis and a
discussion of the various efficiencies that play a role are presented in
section~\ref{sec:analysis}. The determination of the unseparated cross sections
and the separation of these cross section into the four different structure functions 
is described in section~\ref{sec:xsecdet}.
The results are presented in section~\ref{sec:results}. The global
features of the separated cross sections are discussed and a comparison
is made with the results of theoretical calculations. In this discussion
the data from~\cite{ack78,bra76,bra77} are also included.
The paper is concluded with a short summary.

\section{General formalism for exclusive pion electroproduction}
\label{sec:formalism}\nopagebreak[4]

\subsection{Kinematics}
\label{sec:kinematics}\nopagebreak[4]

\begin{figure}[t]
\begin{center}
        \includegraphics[width=9cm]{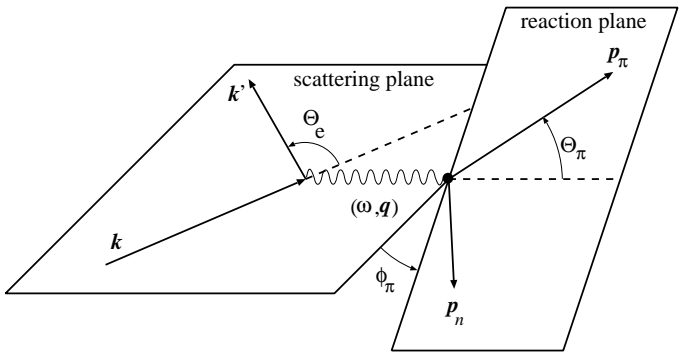}
        \caption{\it (Color online) Kinematics of the \heepi\ reaction in the laboratory frame. 
        }
        \label{fig:eepi_kinematics}
\end{center}
\end{figure}

The kinematics of the \heepi\ reaction are displayed in
Fig.~\ref{fig:eepi_kinematics}. The three-momentum vectors
of the incoming and scattered electrons are denoted by
$\mathbf{k}$ and $\mathbf{k}'$, respectively. Together they define 
the scattering plane. The corresponding four-momenta are
$k \equiv (E,\mathbf{k})$ and $k'\equiv (E',\mathbf{k'})$.
The electron scattering angle is denoted by $\theta_{e}$.
The four-momentum of the transferred virtual photon,
$q \equiv (\omega,\mathbf{q})$, is given by $q \equiv k-k'$.
As usual, the variable \qsq\ is defined as the negative of the transferred 
four-momentum squared: $Q^2 \equiv -q^2$.
The three-momentum $\mathbf{q}$ and the three-momentum vector of the pion
$\mathbf{p}_\pi$ together define the reaction plane.
The angle between the scattering plane and the reaction
plane is denoted by $\phi_\pi$, while the angle (in the lab system)
between $\mathbf{p}_\pi$ and $\mathbf{q}$ is $\theta_\pi$.

The missing energy and missing momentum are defined as:
\begin{eqnarray}
\label{eq:em}
E_m & = & E_e - E_{e'} - E_\pi, \\
\label{eq:pm}
\mathbf{p}_m & = & \mathbf{q} - \mathbf{p}_\pi.
\end{eqnarray}
The missing mass of the recoil system can then be expressed as
$M_m = \sqrt{E^2_m - {\bf p^2}_m}$. In the case of the reaction \heepi\ 
the missing mass is given by the neutron mass $M_m = m_n$.

The \heepi\ reaction can conveniently be described using three
Lorentz~invariants. In addition to $Q^2$, we use the invariant mass
of the virtual photon-nucleon system, $W$, which can be expressed as
$W=\sqrt{M_p^2+2 M_p \omega-Q^2}$, where $M_p$ is the proton mass,
and the Mandelstam variable $t=(p_\pi-q)^2$.
The latter can be expanded into
\begin{equation}
\label{eq:def_t}
t=(E_\pi-\omega)^2 - |\mathbf{p}_\pi|^2 - |\mathbf{q}|^2 + 2 \hspace{1mm}
|\mathbf{p}_\pi|\hspace{1mm}|\mathbf{q}| \cos{\theta_\pi}.
\end{equation}
In the \heepi\ reaction $t$ is always negative. The minimum value 
$-t_{\mathrm{min}}$ of $-t$ is reached for $\theta_\pi=0$.
The minimum value of $-t$ increases for increasing values of
\qsq\ and decreasing values of $W$.

\subsection{Cross sections}
\label{subsec:cross_sections}\nopagebreak[4]

Describing the incoming and outgoing electrons by plane waves,
the cross section for the \heepi\ reaction can be written in the
one-photon exchange approximation as
\begin{equation}
\label{eq:fivefoldxs}
\frac{d^5 \sigma}{dE' d\Omega_{e'} d\Omega_\pi} = \Gamma_V
\frac{d^2 \sigma}{d\Omega_\pi}.
\end{equation}
Here $\Gamma_V$ is the virtual photon flux factor
\begin{equation}
\label{eq:gammav}
\Gamma_v=\frac{\alpha}{2\pi^2} \frac{E'}{E} \frac{K}{Q^2}\frac{1}{1-
\epsilon},
\end{equation}
where $\alpha$ is the fine structure constant, the factor 
$K=(W^2-M_p^2)/(2 M_p)$ is the equivalent real-photon energy, which is the 
laboratory energy a real photon would need to produce a system with 
invariant mass $W$, and
\begin{equation}
\label{eq:epsilon}
\epsilon=\left(1+\frac{2 |\mathbf{q}|^2}{Q^2} \tan^2\frac{\theta_{e}}{2}
\right)^{-1}
\end{equation}
is the polarization of the virtual photon.
The two-fold differential cross section can be
written in terms of an invariant cross section as
\begin{equation}
\label{eq:d2sigma}
\frac{d^2 \sigma}{d\Omega_\pi}= J \frac{d^2 \sigma}{dt d\phi},
\end{equation}
where $J$ is the Jacobian for the transformation from $\Omega_\pi$
to $t,\phi$.

The cross section can be decomposed into four structure functions
corresponding to the polarization states of the virtual photon,
a longitudinal one (L), a transverse one (T), and two interference
terms (LT and TT)~\cite{structurefunctions}:
\begin{eqnarray}
\label{eq:sepsig1}
2\pi \frac{d^2 \sigma}{dt d\phi} & = & 
   \epsilon \hspace{0.5mm} \frac{d\sigma_{\mathrm{L}}}{dt} +
   \frac{d\sigma_{\mathrm{T}}}{dt} + \sqrt{2\epsilon (\epsilon +1)}
   \hspace{1mm}\frac{d\sigma_{\mathrm{LT}}}{dt}
   \cos{\phi}  \nonumber \\
   & & + \epsilon \hspace{0.5mm}
   \frac{d\sigma_{\mathrm{TT}}}{dt} \hspace{0.5mm} \cos{2 \phi} ,
 \end{eqnarray}
where the $d\sigma_\mathrm{X}/dt$ depend on \qsq , $W$ and $t$.
The  dependence of the interference structure functions on $\theta_\pi$
features the following leading order behavior~\cite{sintheta}:
$d\sigma_\mathrm{LT}/dt \sim \sin\theta_\pi$ and
$d\sigma_\mathrm{TT}/dt \sim \sin^2\theta_\pi$.
Therefore the interference structure functions are zero in parallel
kinematics ($\theta_\pi=0$), i.e., at $t_{min}$.

The four structure functions can be isolated if data are taken at
different values of $\epsilon$ and $\phi_\pi$, while $W$, \qsq\ and $t$
are kept constant.
The photon polarization $\epsilon$ can be varied by changing the electron
energy and scattering angle (the so-called ``Rosenbluth-'' or L/T-separation).
The angle $\phi_\pi$ can be varied by measuring the pion left and right of
the $\mathbf{q}$-vector, and out of the scattering plane.

\section{Experiment and Setup}
\label{sec:experiment}\nopagebreak[4]

The two $F_{\pi}$ experiments were carried out in 1997 (Fpi~-~1~\cite{e93021}) and
2003 (Fpi-2~\cite{e01004}) in Hall C at JLab.
The unpolarized electron beam from the CEBAF accelerator was incident on a
liquid hydrogen target. Two moderate acceptance, magnetic focusing spectrometers were 
used to detect the particles of interest. The produced charged pions were detected in the High
Momentum Spectrometer (HMS), while the scattered electrons were detected in the
Short Orbit Spectrometer (SOS).

\subsection{Experiment Kinematics} 
\label{sec:kinexp}\nopagebreak[4]

The choice of kinematics for the two experiments was based on maximizing the
range in $Q^2$ for a value of the invariant mass $W$ above the resonance
region. For each $Q^2$, data were taken at two values of the virtual photon
polarization, $\epsilon$, with $\Delta \epsilon$ typically $>$ 0.25. This allowed for
a separation of the longitudinal, transverse, LT, and TT cross sections.

Constraints on the kinematics were imposed by the maximum available electron
energy, the maximum central momentum of the SOS, and the minimum HMS angle. 
The central kinematics
for the two experiments are given in Table \ref{tab:kinematics}.
In parallel kinematics, \ie, when the pion spectrometer is
situated in the direction of the $\mathbf{q}$ vector, the acceptances of the
two spectrometers do not provide a uniform coverage in $\phi_\pi$.
Thus, to attain full coverage in $\phi_{\pi}$, additional data were
taken with the HMS at a slightly smaller and larger angle compared to the
central angle for the high $\epsilon$ settings. At low $\epsilon$ 
only the larger angle setting was possible.
 \begin{table}[!htb]
 \centering  
 \renewcommand{\arraystretch}{1.2}
 \begin{tabular}{ccccccccc} \hline
 $Q^2$     & $W$    & $|t|_{min}$  & $E$     & $\theta_{e}$  & $E^\prime$   & $\theta_{\pi}$ & $P_{\pi}$  &  \eps\    \\
 (GeV$^2$) & (GeV)  & (GeV$^2$)    & (GeV)   &  (deg)        &  (GeV)       &  (deg)       &  (GeV)     &           \\
\hline 
\multicolumn{9}{|c|}{Fpi-1 Settings} \\
\hline
  0.60     & 1.95   & 0.030        & 2.445   &  38.40        &   0.567      &  9.99
\footnote{Here, the value of $\theta_\pi$ denotes the angle of the momentum transfer $\theta_q$. The 
actual HMS angle was 10.49$\deg$.}
       & 1.856      &  0.37     \\
  0.60     & 1.95   & 0.030        & 3.548   &  18.31        &   1.670      &  14.97       & 1.856      &  0.74     \\
  0.75     & 1.95   & 0.044        & 2.673   &  36.50        &   0.715      &  11.46       & 1.929      &  0.43     \\
  0.75     & 1.95   & 0.044        & 3.548   &  21.01        &   1.590      &  15.45       & 1.929      &  0.70     \\
  1.00     & 1.95   & 0.071        & 2.673   &  47.26        &   0.582      &  10.63       & 2.048      &  0.33     \\
  1.00     & 1.95   & 0.071        & 3.548   &  25.41        &   1.457      &  15.65       & 2.048      &  0.65     \\
  1.60     & 1.95   & 0.150        & 3.005   &  56.49        &   0.594      &  10.49       & 2.326      &  0.27     \\
  1.60     & 1.95   & 0.150        & 4.045   &  28.48        &   1.634      &  16.63       & 2.326      &  0.63     \\
 \hline
\multicolumn{9}{|c|}{Fpi-2 Settings} \\
 \hline
  1.60     & 2.22   & 0.093        & 3.779   &  43.10        &   0.786      &  9.53\footnote{The 
actual HMS angle was 10.50$\deg$.}       & 2.931      &  0.33      \\
  1.60     & 2.22   & 0.093        & 4.709   &  43.10        &   1.650      &  12.54       & 2.931      &  0.58      \\
  2.45     & 2.22   & 0.189        & 4.210   &  51.48        &   0.771      &  9.19\footnote{The 
actual HMS angle was 10.54$\deg$.}       & 3.336      &  0.27      \\
  2.45     & 2.22   & 0.189        & 5.246   &  29.43        &   1.740      &  12.20       & 3.336      &  0.54      \\
 \hline  
 \end{tabular}
 \caption{\label{tab:kinematics} 
\it The central kinematic settings used in the experiments.  In addition,
settings were taken with the pion arm (HMS) at smaller and larger angles
($\theta_\pi$=$\theta_q \pm 4\deg $ in Fpi-1 and $\theta_{\pi}$=$\pm 3\deg $ in Fpi-2)
for the high-$\epsilon$ settings and at the larger angle only for the
low-$\epsilon$ data. The scattered electron was always detected in the SOS.
}
 \end{table}

\subsection{Accelerator} 
\label{sec:accelerator}\nopagebreak[4]

The experiments made use of the unpolarized, continuous wave (CW, 100\% duty
factor) electron beam provided by the JLab accelerator~\cite{ceb93, ceb01}.
The beam has a microstructure that helps in the identification of coincident
events, which is further described in section~\ref{sec:analysis}.
Beam currents were between 10 and 100 $\mu$A.

In order to precisely determine the kinematics, the beam position and angle on
target were monitored using Beam Position Monitors (BPM). The accuracy of the
position measurement was about 0.5~mm and about 0.2~mrad for the incident
angle.
In the Fpi experiments, the beam current was measured by two Beam Current
Monitors (BCM1 and BCM2).
To minimize drifts in the gain, both BCMs are calibrated to an absolute
reference. The calibration is performed using an Unser current
monitor~\cite{uns92}, which has an extremely stable gain, but suffers from
large drifts in the offset on short time scales. The run-to-run uncertainty in
the current as measured by BCM1 and BCM2 was found to be about 0.2\% at 100
$\mu$A. Adding the normalization uncertainty from the Unser monitor, which is estimated
to be 0.4\%, results in an absolute uncertainty for the charge measurement of
0.5\%. A more detailed description of the beam current monitors
can be found in~\cite{nic98}

In order to reduce local density reductions of the liquid targets, the beam was
rastered using a pair of fast raster magnets to a 1.2$\times$1.2 mm$^2$ pattern
during Fpi-1 and to a 2$\times$2 mm$^2$ profile during Fpi-2. The raster position
was recorded event by event. A more detailed description of the fast raster
system can be found in~\cite{yan951}

The energy of the electron beam in Hall C is measured using the deflection of
the electron beam in a known magnetic field in the Hall C arc. Including the
uncertainty in the field integral and the angular uncertainty, the beam energy
can be determined with a precision of $\frac{\delta p}{p}$ $\approx$ $5 \times
10^{-4}$. A detailed description of the beam energy measurement using the arc
method is available in \cite{yan93}.

\subsection{Target}
\label{sec:target}\nopagebreak[4]

The two Fpi experiments used a three-loop cryogenic target stack, mounted
together with a special optics target assembly.  The cryogenic targets
use the same coolant supply and are cooled on the
cryotarget ladder simultaneously.  Two different cryogenic target cell types
were used.  In the Fpi-1 experiment, a 4.5~cm long cylindrical cell with
the axis mounted horizontally and parallel to the beam direction was used
(horizontal flow ``beer can'' design).  In the Fpi-2 experiment, a 4.0~cm
diameter cylindrical cell with vertical axis (vertical flow ``tuna can''
design) was used.  The cell walls are made from Aluminum alloy T6061 with a
thickness of 0.0127~cm (the beer can front wall is half as thick). The
cryogenic targets are typically kept at a nominal operating temperature about
2~K below the boiling point. The hydrogen target was kept at a temperature of
19~K, giving a density of 0.0723 $\pm$ 0.0005 g/$cm^3$~\cite{dun98}.
Cell temperatures were kept constant to within 100~mK during the experiment.
Since the uncertainty in temperature gives a negligible contribution, the
uncertainty in the target density is completely due to the equation of state.
 \begin{figure}[!htb]
 \begin{center}
 \includegraphics[width=3.5in]{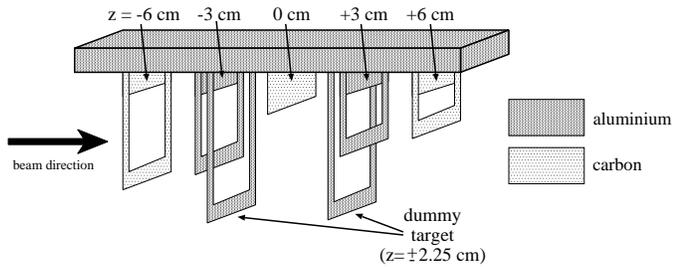}
 \caption{\label{fig:quintar} \it (Color online) Schematic view of the optics target assembly
 ``Quintar'' (not to scale).}
 \end{center}
 \end{figure} 

The optics target assembly was mounted beneath the cryogenic target ladder.  It
consists of five carbon foils (Quintar) and two aluminum foils. A schematic
design of the Quintar is shown in Fig.~\ref{fig:quintar}.  By moving the target
stack vertically, the five targets can be moved into the beam individually or
simultaneously. The solid carbon foils are used with the beam incident on two
or five (``quintar'') foils simultaneously for the purpose of calibrating the
vertex position ($z$) along the beam direction (see section~\ref{sec:optics}). During 
Fpi-1, the Quintar $z$-positions (relative to the nominal target center)
were $z$=$\pm$6.0~cm, $\pm$3.0~cm, 0~cm, while in Fpi-2 they
were $z$=$\pm$7.5~cm, $\pm$3.8~cm, 0~cm. The two aluminum foils
situated at z=$\pm$2.25~cm (Fpi-1) or z=$\pm$2.0~cm (Fpi~-~2) constitute the
``dummy target'' used to measure the contribution of the aluminum cell wall to
the cryotarget yields.  The material of the Aluminum dummy targets is 
Al-T7075 ($\rho$=2.795~g/cm$^2$),
a higher strength alloy. The dummy target foils are approximately seven times
thicker than the cryotarget cell walls.   Further details on the mechanical
aspects of the cryotargets can be found in \cite{dun98, mee98, ter98}.

\subsection{Spectrometers}
\label{sec:spectrometers}\nopagebreak[4]

\begin{figure}[!htb]
 \begin{center}
 \includegraphics[width=3.5in]{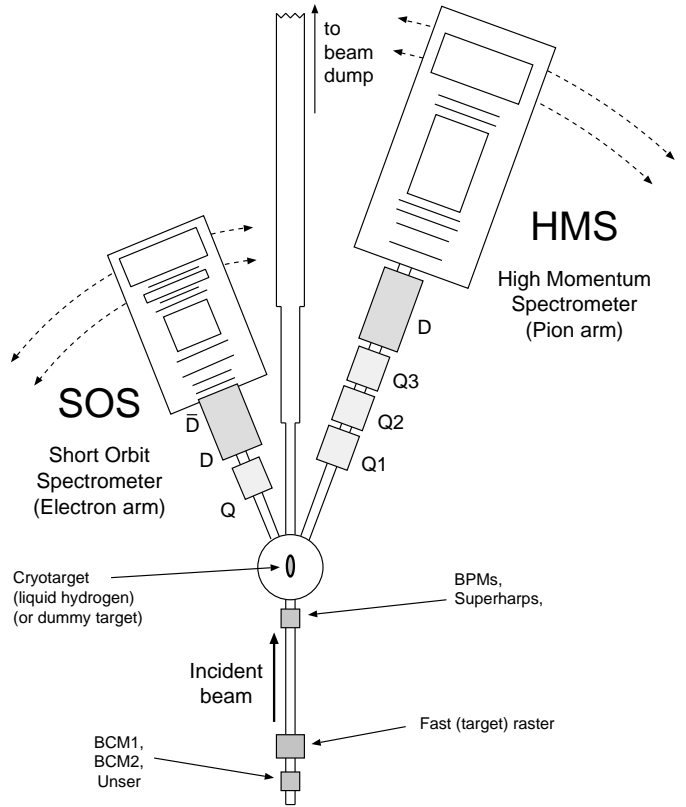}
 \caption{\label{fig_hallc_spectrometers} \it (Color online) Schematic view of the Hall C
 spectrometers with the target and beamline.}
 \end{center}
 \end{figure}

A schematic overhead view of the Hall C spectrometers is shown in
Fig.~\ref{fig_hallc_spectrometers}. Both spectrometers have a relatively
large momentum and solid angle acceptance and are equipped with similar and
highly versatile detector packages. The Short Orbit Spectrometer (SOS), which
was optimized for the detection of short-lived particles, has a relatively
short flight path of about 7.4~m and a maximum central momentum of
1.74~GeV/c.  The High Momentum Spectrometer (HMS) has a 26~m pathlength and a
maximum central momentum of 7.5~GeV/c.

\subsubsection{High Momentum Spectrometer (HMS)}
\label{sec:HMS}\nopagebreak[4]

The HMS consists of three superconducting quadrupole magnets and a
25$^\circ$ vertical-bend dipole magnet used in a point-to-point tune for the
central ray. The momentum acceptance of the HMS is about $\pm$ 10\%.
All magnets are mounted on a common carriage, which can be moved on
rails around a rigidly mounted central bearing. 
A detailed description of the spectrometer hardware is given in
~\cite{arr98}. The design specifications are given in
Table~\ref{tab:HMSspec}.
 \begin{table}[!htb]
     \centering  
     \renewcommand{\arraystretch}{1.2}
\small
     \begin{tabular}{||l|l|l||} \hline       
        Quantity &  HMS & SOS \\
        \hline
        Max. Central Mom. & 7.5~GeV/$c$ & 1.74~GeV/$c$ \\
        Optical Length & 26.0~m & 7.4~m\\
        Angular Range & 10.5$\deg$ to 85$\deg $ & 13.4$\deg$ to 165$\deg$\\
        Momentum Acceptance & $\pm$10\% & $\pm$20\% \\
        Momentum Resolution & $<$0.1\% & 0.1\% \\
        Solid Angle\footnote{The solid angle and angular acceptances 
        are given for the large collimators in both 
        the HMS and SOS spectrometers.} & 6.7~msr & 7.5~msr \\
        In-Plane Ang. Acc.$^a$ & $\pm$27.5~mrad & $\pm$57.5~mrad \\
        Out-of-Plane Ang. Acc.$^a$ & $\pm$70~mrad & $\pm$37.5~mrad \\
        In-Plane Ang. Res. & 1.0~mrad & 2.5~mrad \\
        Out-of-Plane Ang. Res. & 2.0~mrad & 0.5~mrad\\
        Extended Target Acc. & $\pm$7~cm & $\pm$1.5~cm\\
        Vertex Recon. Accuracy & 2~mm & 1~mm\\
        \hline  
     \end{tabular}
     \caption{\label{tab:HMSspec} \it The nominal specifications for the High
     Momentum Spectrometer (HMS) and the Short Orbit Spectrometer (SOS).}
 \end{table}

The HMS detector stack shown in Fig.~\ref{fig_hms_detectors} is situated in a
concrete shielding hut 26~m from the spectrometer pivot.  In order to minimize
multiple scattering and to provide thermal insulation, the region between the
first quadrupole (Q1) and the entrance to the shielding hut is evacuated. The
vacuum region is separated from the surounding environment by vacuum windows.
During Fpi-1, a mylar spectrometer exit window was used; this window was
replaced with a 0.508~mm titanium window (radiation length=3.56cm) prior to the
Fpi-2 experiment.
A detailed discussion of the Hall C spectrometer vacuum system and 
vacuum windows can be found in~\cite{arr98}.

The angular acceptance of the HMS is defined by a collimator positioned in a
collimator box between the target and the first quadrupole magnet.
The collimator box contains two octagonal collimators (''large'' and ''small''),
a sieve slit, which is exclusively used for optics calibration (see 
section~\ref{sec:optics}), and an empty position.
The large collimator, which was used in the experiments, gives a solid angle of 6.8 msr.
The collimators are made from 3.175 cm thick HEAVYMET, which is a
machinable tungsten alloy with 10\% CuNi. The large and small
collimators are flared along the inside edge to match the particle distribution
emanating from the target, but the holes in the sieve slit collimator are not.
The front face of the collimator is at a distance of 166.4~cm from the center
of the target.
A vacuum extension (``snout'') in front of the collimator box limits the amount
of air between the target chamber vacuum and the vacuum inside the HMS to 15~cm.
With this configuration the minimum central angle is about 10.5$\deg$.

In order to set the HMS momentum in a reproducible fashion, the dipole
is set by field using an NMR probe in the magnet, with a reproducibility of the
magnetic field at the level of one part in 10$^4$ and a stability to within one
part in 10$^5$.  The quadrupoles are set by current using a special
procedure to ensure reproducibility~\cite{vol00} and are monitored using the power supply readback
current and Hall probes.

In 1998, the Hall probes indicated a relatively large current offset in
the third quadrupole, which was addressed through a correction to the magnet
field setting routine~\cite{gaskell01}.
During the Fpi-2 experiment, it was found that a small offset in
the third quadrupole (Q3) set current persisted.
This residual Q3
offset was addressed by an ad-hoc correction to the reconstruction of all data
as will be described further in section~\ref{sec:offsets}. In practice, its
influence on the optical properties in the extraction of the final result
is negligible.

\subsubsection{Short Orbit Spectrometer (SOS)}
\label{sec:SOS}\nopagebreak[4]

The SOS spectrometer, which is a {\em QD\={D}} configuration, is a copy
of the Medium Resolution Spectrometer at LAMPF~\cite{lampf}. 
The three magnets are non-superconducting and water cooled, and rest on a common carriage.
The quadrupole focusses in the non-dispersive direction, the first dipole bends
particles with the central momentum upwards by 33$\deg$, and the second one bends them 
downwards by 15$\deg$. In addition to the quadrupole magnet, the
fringe fields arising from the curved shape of the pole tips of the dipole
magnets provide focussing.
A collimator box similar to the one discussed above for the HMS is
attached to the front of the quadrupole.
The specifications of the SOS are given in Table~\ref{tab:HMSspec}.

The SOS magnets are set by field, measured with Hall probes, providing a
short--term reproducibility of $\pm$1.5 Gauss, with long--term drifts of a few
parts in 10$^4$. To ensure that the magnetic fields always lie on the same
hysteresis curve, a particular cycling procedure was used.
At the highest momenta, a correction to the central momentum was applied to
account for saturation effects from the iron of the magnets (see section~\ref{sec:optics}).

\subsection{Detector packages}
\label{sec:detectors}\nopagebreak[4]

\begin{figure}[!htb]
 \begin{center}
 \resizebox{\columnwidth}{!}{\rotatebox{-90}{\includegraphics{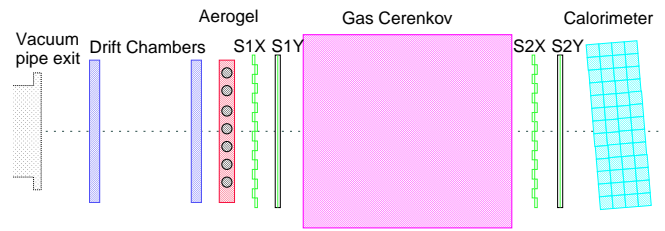}}}
 \caption{\label{fig_hms_detectors} \it (Color online) Schematic sideview of the HMS detectors during Fpi-2. The lead glass
 calorimeter is tilted
 5$^\circ$ relative to the central ray to minimize the loss of
 particles in the spaces between the calorimeter blocks.}
 \end{center}
 \end{figure}

The detector packages in the HMS and SOS are similar and consist of two horizontal drift
chambers for track reconstruction, four scintillator hodoscope arrays used for
 triggering and time-of-flight measurements, and a threshold gas \v{C}erenkov
detector and lead-glass calorimeter for particle identification (mainly
pion-electron separation).
A schematic view of the HMS detector package is shown in
Fig.~\ref{fig_hms_detectors}.
For the Fpi-2 experiment, an aerogel \v{C}erenkov detector (shown in
Fig. \ref{fig_hms_detectors}) was added to the 
HMS detector package to enhance the pion-proton separation at higher momenta.
The individual detector components and their significance for data analysis are
described in the following sections. A complete review of the detector packages
including the detailed geometry and performance evaluation can be found in
\cite{arr98,mee98,nic98,wes99}.

\subsubsection{Drift chambers}
\label{sec:dc}\nopagebreak[4]

Both spectrometers are equipped with a pair of drift chambers. Each 
drift chamber contains six planes of sense wires with a spacing of 1~cm. The
wires are oriented in three (SOS) or four (HMS) different directions 
to allow the measurement of the $x$ and $y$ hit-positions of an incident
charged particle. The redundancy in number of planes
helps to resolve the ambiguity of multiple hits, to determine on
which side of a wire a particle has passed (``left-right ambiguity''), and
to determine a single-chamber estimate of the particle trajectory.

A detailed description of the HMS drift chambers can be found in
\cite{bak95}. The wire planes are ordered {\em x, y, u, v, y', x'}. The {\em x}
and {\em y} planes measure the vertical and horizontal track position,
respectively. The {\em u} and {\em v} plane wires are rotated by $\pm$15$\deg$
with respect to the {\em x} wires. This small angle makes the {\em u} and {\em
v} planes {\em x}-like, with the effect that the redundancy in the {\em x}
direction is good, but poor in the {\em y} direction. The position resolution
for the HMS drift chambers is typically 150~$\mu$m per plane. The two drift
chambers are placed at distances of 40~cm before and after the HMS nominal
focal plane.

The planes in the SOS drift chambers are ordered {\em u, u', x, x', v,
v'}. There are no explicit {\em y} planes, but the {\em u} and {\em v} wire
planes are rotated by $\pm$60$\deg$ with respect to the {\em x} wires. As a
result, the {\em y} resolution of the SOS detector is better than in the
HMS. Unlike in the HMS, the wire planes form pairs with the sense wires offset
by half a cell spacing (0.5~cm). That means that the left-right ambiguity is
resolved if both planes of a pair are hit. The position resolution of the SOS
drift chambers is approximately 200~$\mu$m per plane. The two drift chambers
are placed $\approx$ 25~cm before and after the nominal focal plane of the SOS.

\subsubsection{Hodoscopes}
\label{sec:hodo}\nopagebreak[4]

Hodoscopes consisting of two scintillator planes are located before and after
the gas \v{Cerenkov} counters in both spectrometers.  In the HMS, the first plane of
each hodoscope is segmented into ``paddles'' in the vertical, the second one in the horizontal
direction. In the SOS, the order is reversed. The hodoscopes serve two purposes:
triggering of the data acquisition system, and measuring the particle velocity
using the time-of-flight between the two hodoscope planes.  Each of the
scintillator paddles in the HMS hodoscopes has a thickness of 1.0~cm and a
width of 8~cm, with an overlap of 0.5~cm, while those in the
SOS have thickness 1.0~cm, width 7.5~cm, and an overlap of
0.5~cm. Each scintillator paddle is read out by phototubes at both ends. The signals of all
photomultipliers on each side of the plane are ORed and the signals from the
two sides then are ANDed to form the signals {\bf S1X}, {\bf S1Y}, {\bf S2X} and {\bf S2Y}. The signal
{\bf S1} ({\bf S2}) is the OR of {\bf S1X} with {\bf S1Y} ({\bf S2X} with {\bf S2Y}). The role of the hodoscope
signals in the trigger system is discussed in Sect.~\ref{sec:trigdaq}.

\subsubsection{Gas \v{C}erenkov Detectors}
\label{sec:cerenkov}\nopagebreak[4]

The HMS \v{C}erenkov detector is a cylindrical tank with two parabolic
mirrors at the end and two photomultiplier tubes inside, mounted on the top and
bottom surfaces.  The gas \v{C}erenkov was filled with $C_4F_{10}$ gas at
79~kPa (Fpi-1) or 47~kPa (Fpi-2)
The index of refraction at these pressures is
1.0011 (Fpi-1), 1.00066 (Fpi-2), giving electron thresholds below 10 MeV/c
and pion thresholds of 3.0 (Fpi-1) or 3.8 GeV/c (Fpi-2).
The SOS \v{C}erenkov detector has four mirrors and four phototubes. The
detector is maintained at atmospheric pressure with Freon-12 (CCl$_2$F$_2$),
with a refractive index of 1.00108, yielding a pion threshold of 3 GeV/$c$,
well above the maximum momentum setting of the SOS. A more detailed description
of the \v{C}erenkov detectors can be found in ~\cite{wes99}.

\subsubsection{Lead-glass calorimeter}
\label{sec:calorimeter}\nopagebreak[4]

Each lead glass calorimeter uses 10 cm $\times$ 10 cm $\times$ 70 cm blocks
arranged in four planes and stacked 13 and 11 blocks high in HMS and SOS
respectively. The entire detector is tilted by 5$^\circ$ relative to the
central ray of the spectrometer to minimize losses due to particles passing
through the gaps between the blocks. More detailed information about the
calorimeter system hardware can be found in ~\cite{arr98}.

\subsubsection{HMS Aerogel \v{C}erenkov Detector}

Above momenta of 3 GeV/c, separation of pions and protons in the HMS by 
measuring the particle velocity with the scintillators of the hodoscopes
is not possible in view of the required time-of-flight resolution.
Therefore, an aerogel threshold \v{C}erenkov detector was added to the HMS detector
package in 2003.
Aerogel with a refractive index of n=1.030 was used as the medium,
giving a pion threshold of 0.57 GeV/c and a proton threshold of 3.8 GeV/c,
allowing rejection of protons up to the highest HMS momentum setting used in Fpi-2 of
3.336 GeV/c.  Further details on the design and testing of the HMS aerogel
\v{C}erenkov detector can be found in~\cite{asa05}.

\subsection{Trigger system and data acquisition}
\label{sec:trigdaq}\nopagebreak[4]

In order to keep the event 
rate below the current limit of the data acquisition system ($\approx$ 3~kHz) 
events of interest are selected by the formation of a combination of logic signals 
that indicate when a particlular set of detectors fired. This
combination is used to decide if the event should be recorded, i.e. a pretrigger 
should be formed.

\begin{figure}[!htb]
 \begin{center}
 \resizebox{\columnwidth}{!}{\rotatebox{-90}{\includegraphics{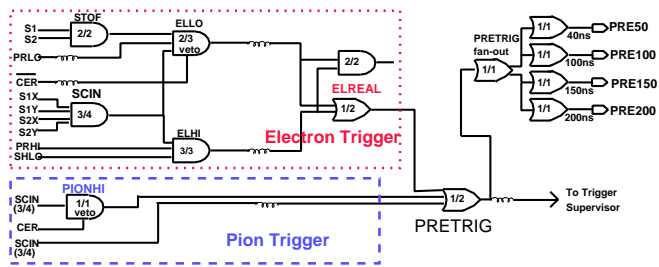}}}
 \caption{\label{fig_pretrigger_logic} \it (Color online) Schematic of the spectrometer pretrigger logic during Fpi-2. The pre-selection of good electron events is accomplished by the upper part of the system, while the lower part is used to select good pion events. The individual triggers from each spectrometer form a pretrigger, which is sent to the trigger supervisor where the signals from the spectrometers are processed and read out of the data is initiated. The split of the {\bf PRETRIG} signal is used to determine the electronic dead time as described in~\ref{subsec:deadtimes}.}
 \end{center}
 \end{figure}

Both spectrometers have a single-arm trigger logic system, which
can be subdivided into two components, one coming from the 
hodoscopes, and one from the combination of signals from the gas
\v{C}erenkov and the calorimeter. The most basic trigger is the {\bf SCIN} trigger
from the hodoscopes, which is satisfied if there was a hit in three out of the
four planes. Fig.~\ref{fig_pretrigger_logic} shows a schematic of the single-arm trigger used. 

The main component of the standard electron trigger, {\bf ELREAL}, is the scintillator 
information, which is provided by two signals ({\bf STOF}, which requires the AND 
of {\bf S1} and {\bf S2}, and {\bf SCIN}). These are used in parallel to give the two 
conditions ({\bf ELHI} and {\bf ELLO}). {\bf ELHI} requires valid scintillator 
information and sufficiently large signals in the calorimeter ({\bf PRHI} and 
{\bf SSHLO}), while {\bf ELLO} is satisfied by 2 out of 3 signals {\bf STOF}, {\bf PRLO}, 
and {\bf SCIN}, and the presence of a signal from the gas \v{C}erenkov.
In the hadron arm, good pion events were selected by {\bf SCIN} with the 
additional requirement of no signal above a given threshold in the \v{C}erenkov ({\bf PIONHI}). 
The threshold for a hit in the \v{C}erenkov was set lower in 
Fpi-1 than in Fpi-2. Each trigger signal is 
sent to a TDC and read out by the data acquisition system. This
makes it possible to determine the efficiency for a given trigger type. The total trigger 
efficiency obtained is discussed in section~\ref{subsect:triggereff}.

In the experiment, the number of pretrigger signals formed for each spectrometer and 
each trigger 
type are recorded. This makes it possible to calculate the computer dead time
for each trigger branch. In addition, the pretrigger signal {\bf PRETRIG} is 
split into four copies of varying length for determining the electronic dead time (see
section~\ref{subsec:deadtimes}). In Fpi-1, the signal was split before
the {\bf PRETRIG} module into four signals of gate widths 30~ns, 60~ns, 90~ns, and
120~ns, whereas in Fpi-2, it was split afterwards
into signals of effective gate widths 50~ns, 100~ns, 150~ns, and 200~ns. More detailed information about the trigger setup can be found in~\cite{moh99,vol00,horn06th}

The data acquisition software used was CODA (CEBAF On-line
Data Acquisition) version 1.4 \cite{abb95}. Three types of data were recorded 
for each run. The TDCs and ADCs for the various detectors were recorded event-by-event,
scalers for e.g. the charge were read out every two seconds, and 
EPICS data from the slow controls were read out at least every 30 seconds (in some cases every 
2 seconds). The ADC, TDC, and scaler information is read out over a network through Fastbus 
and VME crates, each of which had their own Read Out Controller CPU, for each event in the data
stream. Both ADCs and TDCs are sparsified. The threshold values of all ADC channels
are determined from 1000 artificial events created at the beginning of each run.

\section{Data Analysis and Calibrations}
\label{sec:analysis}\nopagebreak[4]

This section describes the determination of the normalized experimental yields, as 
a function of the relevant kinematical variables, including the necessary calibrations, 
with special attention to the precision obtained.

\subsection{Beam position and direction}
\label{sec:beamposdir}\nopagebreak[4]

The position and direction of the electron beam incident on the target were carefully 
monitored during the experiment with the equipment described in section~\ref{sec:accelerator}. 

Deviations in the vertical direction and position of the beam result in
offsets of the momentum and out-of-plane angle of the detected particle,
while deviations in the horizontal direction of the beam result in an
offset in the scattering angle (a deviation in the horizontal position
is taken into account by the optical calibration of the spectrometers).

Observed deviations in the vertical position were 0.3~mm, with a stability of better than 0.2~mm,
and 0.5~mrad in the directions,
with run-to-run variations of less than 0.1~mrad.
Corrections for the effect of these deviations were made 
(see section~\ref{sec:optics}). For example, for a 1~mm vertical offset of the
beam on the target, the reconstructed momentum and 
out-of-plane angle in the HMS would shift by 0.08\% and 1.1~mrad. The corresponding values for 
the SOS are 0.04\% and 0.4~mrad~\cite{horn06th}.

\subsection{Target Thickness}
\label{sec:targetdens}\nopagebreak[4]

As mentioned in section~\ref{sec:target} the nominal target density was 0.0723 $\pm$ 0.0005 g/cm$^3$. 
The effective target length was calculated as the cryotarget length, measured at room temperature, 
corrected for thermal contraction (about 0.4\% at 20~K) of the aluminum cell walls, the offset of 
the cryotarget from the nominal position, and for the central position and the rastering of the beam 
on the target. The latter two corrections were negligible in Fpi-1 thanks to the nearly flat surface of the 
beer-can type cells used. 
In case of Fpi-2, for the largest deviation of the beam from the target center the correction of the 
target length was 1.50 $\pm$ 0.05\%, 
while the corrections for the rastering of the beam were $<$0.1\%.

The effective target length, not corrected for (run dependent) beam offsets, corresponding target thickness, 
and associated uncertainties are listed in Table~\ref{table_cryotarget_thickness}.
The uncertainty on the nominal target thickness was taken as the quadratic sum of a 0.6\% uncertainty on the 
effective target length and the 0.7\% on the target density. The variation in target thickness due to the 
central beam position between high and low $\epsilon$ settings was 0.2\%.
  \begin{table}
 \begin{center}  
  \begin{tabular}{||c|c|c|c||} \hline  
  Experiment & Target  & L$_{target}$       &   t$_{cryogen}$ \\
             &         & (cm)               &   (g/$cm^2$)   \\
  \hline  
   Fpi-1      & LH$_2$  & 4.53  $\pm$ 0.025   &   0.328 $\pm$ 0.003    \\
  \hline  
   Fpi-2      & LH$_2$  & 3.92 $\pm$ 0.025   &   0.283 $\pm$ 0.003    \\
  \hline  
    \end{tabular}
  \end{center}
\caption{\label{table_cryotarget_thickness} \it The cryotarget lengths and thicknesses, not corrected for beam offsets.}
 \end{table}

Although the electron beam was rastered to spread the energy deposited in the target 
liquid over a larger volume, the 
target thickness may still be influenced by local target boiling. 
To measure the effective target thickness, 
$^1$H$(e,e)$ elastic 
scattering data were taken at fixed kinematics for electron beam currents between 10 and 
90~$\mu$A. A possible target thickness reduction was determined by comparing the deadtime- 
and tracking-corrected yields as a function of beam current. To check that rate-dependent 
effects were properly taken into account, additional data were taken with a solid carbon 
target during Fpi-2, for which no density reduction effects are expected.  
The results suggest no current and/or rate dependent effects for carbon at the $10^{-3}$ 
level.
For the cryogenic hydrogen target, the analysis of Fpi-1 data taken with the horizontal-flow 
cryotarget and a 
fast raster amplitude of $\pm$1.2~mm, gave a yield reduction of (6$\pm$1)\%/100$\mu$A. The 
Fpi-2 yield reduction for the vertical-flow cryotarget was determined to be (0.6$\pm$0.1)\%/100$\mu$A for 
a raster amplitude of $\pm$2~mm. The improvement in the yield reduction in Fpi-2 compared to Fpi-1 is due 
to the improved raster design and vertical-flow cryotarget.

\subsection{Optical calibrations}
\label{sec:optics}\nopagebreak[4]

The HMS and SOS spectrometers were used to determine the momentum vector (magnitude and direction) 
of the detected particles at the target, as well as to reconstruct the location of the reaction vertex.
The reconstruction of the vertex kinematics is achieved by means of a matrix containing the elements 
of a Taylor-expansion of the vertex variables in terms of the focal-plane variables.
These variables, which are determined from the drift chamber information, are the positions 
$x_\mathrm{fp}$ in the dispersive and $y_\mathrm{fp}$ in the non-dispersive direction, and the 
directions $x'_\mathrm{fp}$ and $y'_\mathrm{fp}$ with respect to the forward $z$-direction, of 
the particle in the detection (or nominal focal) plane. This plane is $\approx$ half-way 
between the two drift chambers (see Sec.~\ref{sec:dc}).
Both spectrometers feature a point-to-point focus in both the dispersive and non-dispersive 
directions for particles with a central momentum, which is the momentum of a particle that passes 
through the middle of the entrance quadrupole(s) and the wire chambers (the optical axis) of the 
spectrometer.
The central momentum, $p_0$, is related to the magnetic field of the spectrometer by $p_0=\Gamma \cdot B$. 
The value of the spectrometer constant, $\Gamma$, is given by the spectrometer design, adjusted on 
account of calibrations.

The reconstruction is performed with the following formula:
\begin{equation}
\label{eq:taylor}
   x^i_{tar} = \Sigma_{j,k,l,m}^N M^i_{jklm}(x_{fp})^j (y_{fp})^k 
(x^\prime_{fp})^l (y^\prime_{fp})^m,
\end{equation}
where the $M^i_{jklm}$ denote the elements of the reconstruction matrix. The reconstructed 
quantities, $x^i_{tar}$ in the target system, are the sideways position $y_\mathrm{tar}$ in 
a plane perpendicular to the optical axis at the target, the inclinations $x'_\mathrm{tar}$ 
and $y'_\mathrm{tar}$ with respect to the optical axis, and the momentum $p$ of the particle. 
The latter is commonly described relative to the central momentum $p_0$ by using the variable 
$\delta$:
\begin{equation}
\delta = \frac{\left(p - p_0\right)}{p_0}.
\end{equation}
The sum over indices is constrained by $0 \leq j + k + l + m \leq N$, where $N$ is the order of the 
series expansion. In the reconstruction it is assumed that $x_{tar}=0$ and that the vertical 
spread of the beam at the target can be neglected, which enables one to determine $\delta$ 
(and thus $p$). Any deviation of $x_{tar}$ from zero, e.g. from rastering the beam, is 
corrected for using the known optical properties of the spectometer. 
The left-right symmetry of the spectrometers restricts the allowed combinations of $k$ and $l$. 
For instance, it forces the matrix elements for $\delta$ and $x'_\mathrm{tar}$ to be zero when $k+l$ 
is odd, while those for $y_\mathrm{tar}$ and $y'_\mathrm{tar}$ are zero when $k+l$ is even. If the 
symmetry is broken, e.g. due to a misalignment of a magnet, the ``forbidden'' matrix elements may 
have non-zero values.

The reconstruction matrix elements were fitted by using specially taken calibration data. For 
determining the $y_\mathrm{tar}$, $x'_\mathrm{tar}$ and $y'_\mathrm{tar}$ matrix elements data 
were taken using the quintar and sieve slits (see section \ref{sec:target}
and \ref{sec:HMS}).
These slits consist of 3.175~cm thick tungsten plates with holes at regular intervals, providing 
for discrete values of $x'_\mathrm{tar}$ and  $y'_\mathrm{tar}$.
The quintar gave discrete values of $z_\mathrm{tar}$, from which the value of $y_\mathrm{tar}$ 
can be calculated by using the angle between the target and the spectrometer, and the value of 
$y'_\mathrm{tar}$. These data were taken with a continuous particle-momentum spectrum. 
Discrete momenta for determining the $\delta$ matrix elements were obtained by using (in)elastic 
scattering data on $^{12}$C and $^1$H targets. By changing the central momentum (or the spectrometer 
angle in case of the hydrogen target) in discrete steps, the scattered-electron peaks were shifted 
over the focal plane, thus scanning the entire $\delta$-acceptance.
\begin{figure}[!htb]
\begin{center}        
\includegraphics[width=2.5in]{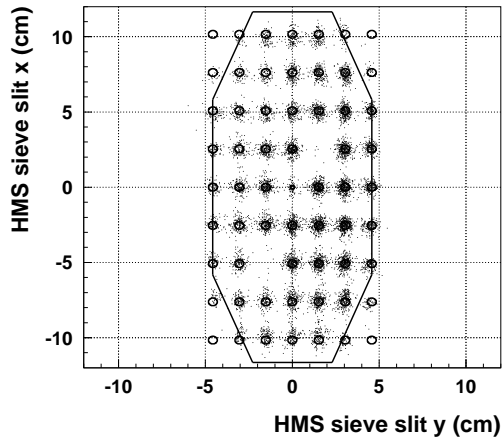}
\caption{\it (Color online) Reconstruction of the hole pattern of the HMS sieve slit. The central hole is smaller, and some holes are blocked for verifying the orientation. Overlayed is the acceptance as defined by the octagonal collimator. The lack of events in the holes in the corners is caused by limited acceptance. Data from all five quintar positions were added.}
\label{fig:hms_sieve_pattern}
\end{center}
\end{figure}
\begin{figure}[!htb]
\begin{center}
        \includegraphics[width=2.5in]{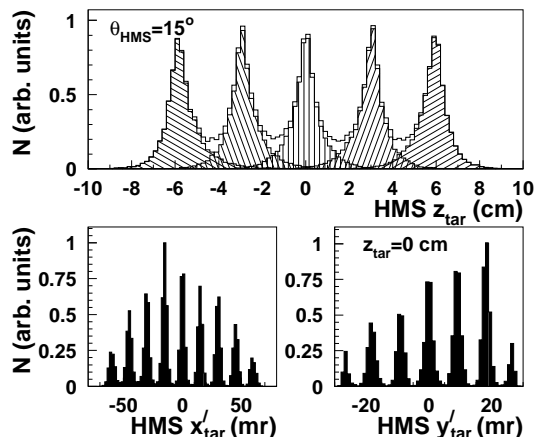}
        \caption{\it (Color online) HMS quintar and sieve-slit reconstruction.
        Top: reconstruction of the quintar $z_\mathrm{tar}$ co-ordinate.
	  The distribution shown is the sum of the five individual targets.
        Bottom: reconstruction of the vertical (left) and horizontal (right) 
        sieve-slit hole patterns (central target foil only).
        }
\label{fig:hms_y_and_holes}
\end{center}
\end{figure}

The strengths of the quadrupole fields for a particular field of the dipole magnet 
(central momentum setting) are selected to obtain point-to-point focussing in both 
directions for particles travelling along the optical axis ($p$=$p_0$, $\delta$=0).
 In this case, the focus of the beam envelope in the focal plane will be located at 
$x_\mathrm{fp}$=0 and $y_\mathrm{fp}$=0. Changes in the magnetic field strength due 
to saturation effects would manifest in a shift of the focal plane focus. The stability 
of the focal plane distributions for the HMS was found to be better than $\pm$0.5 cm 
for central-momentum settings ranging from 0.8 to 5.0~GeV/c.

The HMS reconstruction matrix was expanded up to fifth order in the 
fitting. Forbidden matrix elements were included, which improved the 
reconstruction, especially for $y_\mathrm{tar}$ and $y'_\mathrm{tar}$
\footnote{It was later found \cite{vol00} that the breaking of midplane symmetry 
(which leads to forbidden matrix elements) is most likely caused by a 
rotation of Q2 by 0.2$\deg$ around its optical axis.
No explicit correction for the effects of this are needed, since the
forbidden matrix elements are included in the model of the HMS.}.
Fig.~\ref{fig:hms_sieve_pattern} displays the sieve slit reconstruction of 
the HMS, overlaid with the nominal hole positions and the area covered by the collimator.
The outermost vertical sieve slit holes are at $\pm$60.5~mrad so that the sieve slit 
does not entirely cover the acceptance of the octagonal 
collimator. For particles passing the octagonal collimator beyond this  
range, the reconstruction relies on the extrapolation of the Taylor 
series (Eq.~\ref{eq:taylor}) to a region where it has not been fitted, 
and the resolution worsens considerably. Therefore, only a range of 
$\pm$60~mrad in $x'_\mathrm{tar}$ was used during the analysis of the 
$\pi^+$ data in Fpi-1. To extend the valid region of the out of plane 
matrix elements, optics data were taken in 2003 with the sieve 
slit shifted by one half row extending the vertical range of the 
outermost sieve hole columns 
by $\pm$ 1.27~cm. The $x^\prime_{tar}$ matrix elements were then 
optimized following the procedure outlined in~\cite{horn06th}. 
In this analysis the HMS reconstruction matrix was expanded to sixth order. 

Fig.~\ref{fig:hms_y_and_holes} shows the reconstruction of the 
$z_\mathrm{tar}$ position of the five quintar target foils, and the 
reconstruction of the sieve slit holes in the vertical ($x'$) and the
horizontal direction ($y'$). 
The resolutions in $x'_\mathrm{tar}$ and $y'_\mathrm{tar}$ were 
determined by quadratically subtracting the $\sigma$ of the shape of 
the holes from the values given above. The resolutions are summarized 
in Table~\ref{tab:opresol}. 
\begin{figure}[!htb]
\begin{center}
        \includegraphics[width=2.5in]{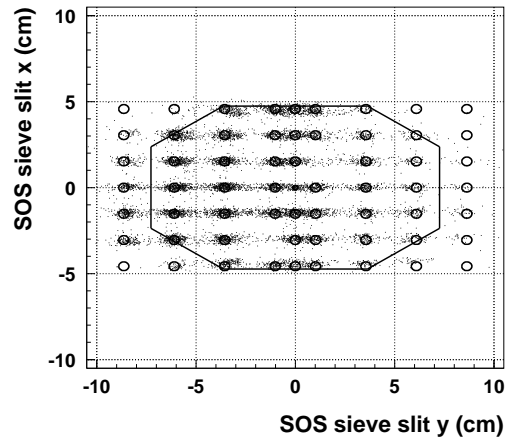}
        \caption{\it (Color online) Reconstruction of the hole pattern of the SOS sieve slit, overlayed
        with the acceptance as defined by the octagonal collimator. The central
        hole is smaller, and some holes are blocked for verifying the
        orientation. The lack of events in the holes in the
        corners is caused by limited acceptance. 
        }
\label{fig:sos_sieve_pattern}
\end{center}
\end{figure}
\begin{figure}[!htb]
\begin{center}
        \includegraphics[width=2.5in]{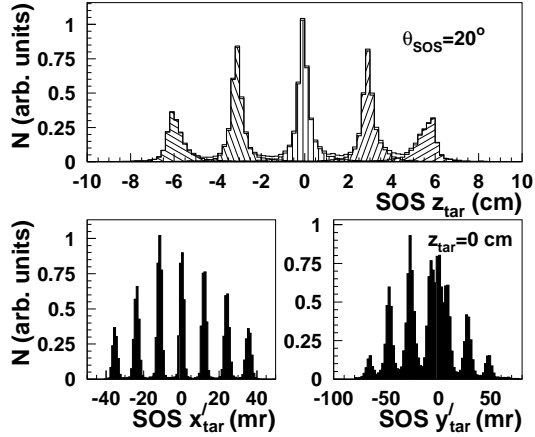}
        \caption{\it (Color online) SOS quintar and sieve-slit reconstruction.
        Top: reconstruction of the quintar $z_\mathrm{tar}$ co-ordinate. The
        distribution shown is the sum of the five individual targets (hashed).
        Bottom: reconstruction of the vertical (left) and horizontal (right)
        sieve-slit hole pattern.
        }
\label{fig:sos_y_and_holes}
\end{center}
\end{figure}

The SOS reconstruction matrix was expanded to sixth order.
The matrix was first determined in 1997 using optics data taken at
P$_{SOS}$ $\approx$ 1.4 GeV/c for $\delta$ and
P$_{SOS}$ $\approx$ 1.65 GeV/c for the quintar/sieve slit data.
The reconstruction of the SOS sieve slit is shown in Fig.~\ref{fig:sos_sieve_pattern}.
The top plot in Fig.~\ref{fig:sos_y_and_holes} shows the reconstruction of the positions of
the target foils of the quintar target with the SOS positioned at an angle
of 20$\deg$ with respect to the beam.
The bottom plot shows the sieve slit pattern for the central foil of the quintar target.
The resolutions are listed in Table~\ref{tab:opresol}.

\begin{figure}[!htb]
 \begin{center}
\includegraphics[width=2.5in]{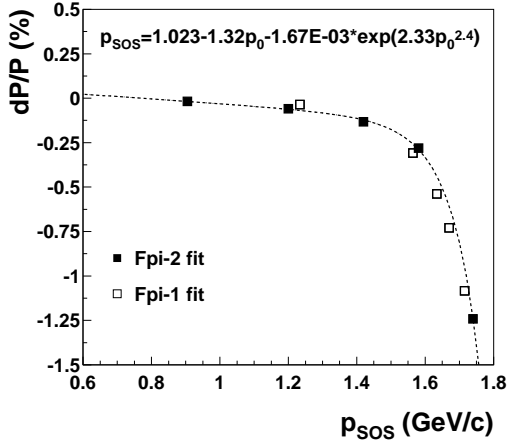}
 \caption{\label{fig:SOS_satcorcurve} \it (Color online) Saturation correction for the SOS central momentum. The 2004 data points are from~\cite{Cla04pc}.}
 \end{center}
 \end{figure}

Complications arise due to the resistive nature of the SOS magnets.
It was found \cite{vol00,gaskell01} that saturation effects start to
play a role for central momentum settings above about 1.0~GeV/c.
The effective field length decreases, resulting in a decrease
of $p_0/B = \Gamma$. A correction to the central momentum was parametrized
based on elastic scattering data from hydrogen, see fig.~\ref{fig:SOS_satcorcurve}.
The effect can be as large as 1.3\% at the maximum central momentum
of 1.74 GeV/c.

A second effect of saturation is that it influences the SOS optics.
This effect was first observed in Fpi-1 and was addressed with a momentum dependent
correction to $\delta$ only, as described in detail in chapter 4.6 of~\cite{vol00}. 
It was addressed in much more detail in Fpi-2 by re-fitting the optics matrix
at different central momenta, thus making the matrix momentum dependent~\cite{xu04}. The 
main effect was on the determination of $\delta$.
The effects on $x'_\mathrm{tar}$ and $y'_\mathrm{tar}$ were found to be relatively
small, of the order of 1 mrad, as were those on $y'_\mathrm{tar}$.
\begin{table}[t]
\begin{center}
\begin{tabular}{|l|l|l|}
\hline
 & HMS & SOS \\
\hline
$x'_\mathrm{tar}$ (indiv. holes) & 1.8~mrad     & 0.3-0.5~mrad \\
$x'_\mathrm{tar}$ (columns)      & 1.8-2.1~mrad & 0.3-0.8~mrad \\ \hline
$y'_\mathrm{tar}$ (indiv. holes) & 0.3-0.7~mrad & 2.4-2.7~mrad \\
$y'_\mathrm{tar}$ (rows)         & 0.8-1.0~mrad & 3.1-3.3~mrad \\ \hline
 $y_\mathrm{tar}$ (mean)         & 2~mm         & 0.9-1.1~mm \\ 
\hline
\end{tabular}
\caption{\label{tab:opresol}
        Resolutions ($\sigma$) of HMS at 2.2~GeV/$c$ and SOS at 
1.65~GeV/$c$. 
        The resolutions $x'_\mathrm{tar}$ and $y'_\mathrm{tar}$ are 
shown for 
        individual holes and for rows and columns of holes in order to 
provide 
        information about the size of systematic effects in the sieve 
slit 
        reconstruction.
}
\end{center}
\end{table}

\subsection{Offsets}
\label{sec:offsets}\nopagebreak[4]

After the optimization of the matrix elements as described in
section~\ref{sec:optics}, the spectrometer quantities 
$\delta$, $x'_\mathrm{tar}$, $y_\mathrm{tar}$ and $y'_\mathrm{tar}$ should be 
reconstructed correctly. However, during the experiment one should allow for
small deviations from the calibration values, resulting, e.g., from
small variations in the vertical position of the beam and in cycling
the spectrometers, and possible saturation effects. Furthermore, small
deviations in the electron energy $E_e$ and the central spectrometer
angles $\theta_{\mathrm{HMS}}$ and $\theta_{\mathrm{SOS}}$ from the nominal
values are possible. 
Most of these experimental offsets can be traced by analyzing single-arm elastic
scattering and coincident $^1$H$(e,e'p)$ data. This reaction is kinematically
overdetermined, which allows one to inspect the following quantities:

\begin{itemize}
\item   the invariant mass of the photon-target system, $W$, which 
should equal the proton mass,
\item   the missing energy, $E_m=E_e-E_{e'}-T_p$, where $E_e$ is the 
        energy of the incoming electron, $E_{e'}$ the energy of the 
	scattered electron, and $T_p$ the kinetic energy of the recoiling proton,
        which should should be zero,
\item   the three components $p_m^{par}$, $p_m^{per}$ and
        $p_m^{oop}$ of the missing momentum $\mathbf{p}_m=\mathbf{p}_e
        - \mathbf{p}_{e'} - \mathbf{p}_p$ (defined as the 
        components parallel to the momentum transfer vector 
        $\mathbf{q}=\mathbf{p}_e - \mathbf{p}_{e'}$, perpendicular to 
        $\mathbf{q}$ in the scattering plane, and out of the scattering 
        plane), which should all be zero.
\end{itemize}

The seven experimental quantities that are checked are the beam 
energy $E$, the momenta of the scattered electron and the recoiling 
proton $p_{e'}$ and $p_{p}$, their angles $\theta_{e'}$ and $\theta_p$, 
and their out-of-plane angles $\phi_{e'}$ and $\phi_{p}$.
The quantities $\phi_{e'}$ and $\phi_{p}$ are related to $p_m^{oop}$,
while the other ones are related to the four quantities $W$, 
$E_m$, $p_m^{par}$ and $p_m^{per}$. 

During the experiment, single arm $^1$H$(e,e^\prime)$ and coincidence
$^1$H$(e,e'p)$ runs were taken at each electron energy.
These data were analysed to yield a set of experimental offsets 
that minimizes the deviations of the values of  $W$, $E_m$ and 
$\mathbf{p}_m$ from their theoretical values. In the analysis, the 
offset in a spectrometer angle was taken to be constant, independent
of the spectrometer setting.
During Fpi-1, the offset in the beam energy, $E$, with respect to the value 
determined as described in Section~\ref{sec:accelerator},
was allowed to be different for each new electron energy.
In view of the availibility of more precise beam-energy measurements
the beam energy was kept fixed during Fpi-2.
Since no saturation effects have been observed in the HMS up to momentum settings of 
5 GeV/c, the offset in the HMS spectrometer momentum was 
taken to be constant for all 
excitations. In the case of the SOS the offset was taken 
to be a function of the central momentum (see subsection~\ref{sec:optics}). 
In fitting the offsets, the effects of radiation and energy loss were taken 
into account. The effect of the beam not being centered 
vertically was included as well, because such an offset can mimic a 
momentum offset. 

The experimental offsets found are listed in Table~\ref{table-kinmat-offsets}.
The major offsets are those on the spectrometer momenta.
With these offsets, the reconstructed values of $W$, $E_m$ and $\mathbf{p}_m$
were within 1-2 MeV or MeV/c of their physical values.
The intrinsic uncertainties (not including possible correlations between 
the offsets) in the offsets are $\pm$ 0.05\% for energies and momenta,
and $\pm$ 0.5~mrad for angles.
\newline
The offsets on the electron energies are $<$ 0.15\%, and the offsets on the  
in-plane spectrometer angles are $<$ 1~mrad.
The larger values of the out-of-plane angle offsets have a few origins.
First of all it is known from surveys that the SOS has a 2.6 mrad out-of-plane offset.
Furthermore, it was determined afterwards that the original calibration data for
both the HMS and SOS had been taken with a vertical offset of the beam.
This influences especially the $\phi$-offset of HMS.
During Fpi-1 no corrections were made for a vertical offset of the beam
during the data taking, but the effect was accounted for in the
$\phi$-offsets of both the HMS and SOS.
During Fpi-2, such corrections were included. The remaining $\phi$-offsets
of 1.1 and 0.6~mrad mainly result from the mentioned offset during the
original calibrations.

The Fpi-2 offsets include no offset in the HMS central angle (compared to
the previously used angle offset of 1~mrad). The Fpi-2 HMS kinematic offsets
are in  relatively good  agreement with elastic electron singles data
from 1999~\cite{chr04} and with data taken in 2004~\cite{Cla04pc}.
The difference in the values found for Fpi-1 and Fpi-2 may partly be due
to a small difference in the direction of the incoming beam. Also for the HMS,
which was used to detect the scattered electron,
there is a strong correlation between the offsets found for $\theta$
and for $p_0$. When using the Fpi-2 offsets for the data taken during
Fpi-1, an only slightly worse description is found
\footnote{Checks have shown that because of the correlation in these offsets,
the uncertainty in them has an almost negligible influence on
the final \heepi\ results, see section~\ref{sec:errorpropagation}.}.
As described in subsection~\ref{sec:optics}, the large momentum-offset
values for SOS result from saturation effects.

  \begin{table}
  \begin{center}  
  \begin{tabular}{|lcc|}
  \hline  
 Quantity  & HMS                  &   SOS            \\
           & Fpi-1 (Fpi-2)          & Fpi-1 (Fpi-2)      \\
 \hline
  $\theta$  &  +1.0  (0.0)  mrad   & -0.4 (0.0)  mrad  \\
  $\phi$    &  +2.4  (+1.1) mrad   & +2.6 (+3.2) mrad  \\
  $p_0$     &  -0.33 (-0.13) \%    & 0.0 to -1.1 (0.0 to -1.4)\% \\
  \hline  
  $E_e$     &  -0.15 to +0.14\% (0.0)          &            \\
  \hline  
  \end{tabular}
  \end{center}
 \caption{\label{table-kinmat-offsets} \it Kinematic offsets. See text for discussion.}
 \end{table}

\subsection{Particle Identification and Event Selection}
\label{sec:pid}\nopagebreak[4]

Electrons were identified in the SOS using the gas \v{C}erenkov and calorimeter. 
Electron events were selected with a \v{C}erenkov cut of N$_{photoelectrons}$ $>$ 0.5 
and a calorimeter cut $E/p >$ 0.6 (Fpi-1) or 0.7 (Fpi-2). The relatively low photoelectron 
cut used resulted in several $\pi^-$ passing particle identification. However, when 
combining these with a $\pi^+$ in the HMS almost all of them were random coincidences and 
were removed by the random subtraction. The loss of electrons due to these cuts was $<0.1\%$, 
whereas the on-line suppression of pions 
was better than 99\%. After off-line analysis the pion 
contamination was $<0.03\%$ in all cases.

In the HMS, where $\pi^+$ were detected, the contaminating particles were protons and positrons. 
During Fpi-1, an upper limit of 0.2 photo-electrons in the \v{C}erenkov detector provided a positron 
rejection of $>99.4\%$, resulting in a final positron contamination of $<0.02\%$. The loss of pions 
at this limit was 3.1\%. Proton rejection was accomplished via 
the particle speed, $\beta=v/c$, calculated from the time-of-flight difference between the 
two hodoscopes in the HMS detector stack. With the chosen cut of $\beta>$0.925, the loss of pions 
is negligible.

During Fpi-2, no off-line \v{C}erenkov detector cuts were applied to eliminate positrons
as those that pass particle identification cuts are removed by the subtraction of random 
coincidences in the analysis (see section~\ref{sec:backgrounds}).
During Fpi-2 the pion and proton momenta were high, resulting in $\beta$ distributions for pions 
and protons which were not completely separated, and the HMS aerogel \v{C}erenkov was used to provide 
additional discrimination. The aerogel \v{C}erenkov efficiency was determined from $\pi^-$ production 
data with tight cuts on the missing mass and the calorimeter to eliminate electrons, and was found 
to be 99.5$\pm$0.02\% for a
threshold cut of N$_{photoelectrons}>3$ (the mean number of photoelectrons
being 12).

Protons passing the particle identification cuts were effectively removed by the subtraction 
of random coincidences. Real proton coincidences were avoided via coincidence time cuts 
(Sec. \ref{sec:backgrounds}). 

\subsection{Efficiencies}
\label{sec:efficiencies}\nopagebreak[4]

In calculating the normalized yield, one must apply corrections for inefficiencies resulting, e.g., 
from track reconstruction and data acquisition deadtime. Various efficiencies are discussed in 
detail in the sections below.
                                                                                    
\subsubsection{Tracking efficiency}
\label{sec:tracking_eff}\nopagebreak[4]

As described in section~\ref{sec:optics}, the basis of kinematic reconstruction is to find a valid 
track in the pair of drift chambers in each spectrometer. 
Each chamber has six planes of wires and a signal in at least five planes is required by the tracking 
algorithm to start constructing a track for a given event. The tracking algorithm performs a $\chi^2$ 
minimization by fitting a straight line through both chambers. 
In case the fit results in more than one possible track, the track that comes
closest to the scintillator paddle in the second hodoscope that fired, 
is selected (this feature was not yet implemented during Fpi-1). The complete hierarchy of selection criteria is described in detail in~\cite{tva04}. Projecting the fitted track to the nominal focal plane 
yields the position ($x_{fp}$,$y_{fp}$) and direction ($x^\prime_{fp}$,$y^\prime_{fp}$) of the particle. 

The tracking efficiency is defined as the probability that the tracking algorithm found a valid track for 
a particle identified as an electron (or pion). It depends on both the efficiency of the wire chambers and 
on the tracking algorithm. The particle identification requirements eliminate the bias introduced by the 
presence of other particle types in the acceptance with possible intrinsic lower efficiency.
The HMS tracking efficiency was generally above 98\% (Fpi-1) or 97\% (Fpi-2) and was only weakly dependent 
on the event rate.
During Fpi-1 the SOS tracking efficiency was slightly worse, but still generally
above 96\%, while during Fpi-2 it was about 99\%.
This improvement is largely due to the improved tracking algorithm used.
The difference between HMS and SOS mainly reflects the difference in incident count rates.

At high rates there is a nonzero probability for more than one particle to pass through
the drift chambers within the approximately 200~ns TDC window used in the analysis. The 
tracking algorithm determines only one, "best", track for each event. Any additional 
tracks are accounted for by either the electronic or computer dead time corrections. However, 
it has been observed that the efficiency for finding a single track is actually significantly
lower in the presence of multiple real tracks (this is due to sofware limitations when 
dealing with many hits). The rate dependence of the tracking efficiency then mostly comes about
from the increased probability to have multiple tracks at high rates. To resolve this issue, a 
tracking efficiency calculation, including multiple track events, was developed 
(see ~\cite{hornTN07} for details).

\subsubsection{Trigger Efficiency}
\label{subsect:triggereff}

The trigger (see section~\ref{sec:trigdaq}) used for pions in the HMS is largely determined by 
the scintillators (plus absence of the \v{C}erenkov signal), so the trigger efficiency can be 
expressed directly in terms of the efficiency of the separate scintillator signals. For SOS, the 
total trigger efficiency is given by the product of scintillator, calorimeter and gas
\v{C}erenkov efficiencies.

The needed 3 out of 4 scintillator efficiency for either spectrometer can be written as:
\begin{eqnarray}
\label{eqn_trig}
   P_{\frac{3}{4}} & = & P_1 P_2 P_3 P_4+P_1 P_2 P_3 (1 - P_4)+P_1 P_2 
(1 - P_3) P_4   \nonumber \\
           & + & P_1 (1 - P_2) P_3 P_4 + (1 - P_1) P_2 P_3 P_4 ,
\end{eqnarray}
where $P_i$ denotes the single-plane efficiency for each scintillator plane.
The individual plane efficiencies can be calculated from the number of times 
a valid track that gives a valid hit in three planes, produces a signal in the 
paddle of the fourth plane it intersects. To minimize the track dependence of 
the efficiency, adjacent paddles to the one that should have fired are included 
in the calculation. 

The variation of the 3/4 efficiency across the spectrometer acceptance is also 
of great importance, since different parts of the acceptance feed different 
parts 
of the phase space. This was investigated for HMS during Fpi-2.
The 3/4 efficiency was determined for both HMS $e-p$ elastic and pion 
electroproduction data. In both cases, an inefficiency of 1.5\% was found at 
negative fractional momentum, $\delta <$ -5.0\% (-8.0\%), outside the region 
used in the analysis of the Fpi data. Within that region the efficiency was 
99.85 $\pm$ 0.05\%.

\subsubsection{Computer and Electronics dead times}
\label{subsec:deadtimes}\nopagebreak[4]

The computer dead time can be directly calculated from the number of (generated) 
pretriggers and (accepted) triggers. 
The computer 
dead time was relatively large during Fpi-1 because the data acquisition system 
was used in an unbuffered mode to avoid potentially serious synchronization problems. 
The event rate was commonly chosen such that computer dead time was below 40\%. 
The computer dead time during Fpi-2 was about 10\%.

The computer dead time at high rates was tested using data taken at fixed current 
and varying computer dead times. The resulting normalized and corrected yields at 
different live times agreed within 0.2\%, which number was taken as the uncertainty 
in the computer live time.

While the computer dead time can be directly measured, the electronic dead time was 
estimated from copies of the original pretrigger signal at varying limiting gate 
widths. This was done using four scalers with different gate widths 
(30~ns, 60~ns, 
90~ns and 120~ns in case of Fpi-1, and 40~ns, 100~ns, 150~ns and 200~ns for Fpi-2). 
The true limiting gate width in the trigger logic corresponds to the width of the 
pretrigger output and was effectively about 50~ns for Fpi-1 and 60~ns for Fpi-2. 
Knowing the rates and the length of the gates of the four scalers, the effective 
limiting gate width, $\tau$, can be determined, and hence the correction for 
electronic dead time, using the formula $\epsilon_{el.d.t.} = 1 - R \tau$, where $R$ 
is the actual event rate. The corrections were at most 3\% (for high rates in the HMS during Fpi-2), 
with an overall uncertainty, calculated from an estimated uncertainty of the gate 
widths of HMS and SOS, of 0.1\%.

\subsubsection{Coincidence Blocking}
{\label{sect:coinblock}}

The coincidence time between the spectrometers is used in the analysis to define good 
coincidence events. Such a coincidence event will normally be started at the TDC with 
a delayed HMS trigger and stopped by the SOS. However, due to interference between 
non-coincident and coincident events, a fraction of events are recorded with a value 
of coincidence time outside the main timing window as defined by the pretrigger signal 
widths. These ``coincidence blocking'' events will be lost from the data due to the 
coincidence time cuts used in the analysis. The coincidence blocking correction was 
estimated from the rate dependence of the number of blocked events. The values range 
from 99.5 to 99.9\% with an uncertainty of about 0.1\%.

\subsubsection{Pion absorption and beta efficiency}
\label{sec:absorption}\nopagebreak[4]

A fraction of the produced pions are lost due to nuclear interactions in the 
materials that they traverse before reaching the detectors in the HMS detector 
hut. The loss is mainly due to absorption and large-angle scattering.

Since the absorption cross sections for protons and pions are rather similar 
for momenta around 2 GeV/c, in Fpi-1, the absorption was estimated based on 
the difference in yield for simultaneously measured $^1$H($e,e^\prime$) and 
$^1$H($e,e^\prime p$) reactions, yielding a value of $4.0\pm 1.5$\%. In Fpi-2, 
the transmission of pions through the spectrometer was calculated using the 
list of traversed material and the pion-nucleon reaction cross section, which 
includes absorption and inelastic reactions. The calculated transmission for 
pions with momenta of 2.93 GeV/c and 3.34 GeV/c was 95\%, with an estimated 
uncertainty of 2\%. The reduced pion transmission compared to Fpi-1 is mainly 
due to the thicker (titanium) spectrometer exit window and the addition of the 
aerogel Cerenkov in the detector stack.

The situation is complicated by the following. In the analysis a cut is used on
 $\beta$-$\beta_p$, where $\beta$ is the particle velocity determined from the 
time of flight between the two scintillator hodoscopes, and $\beta_p$ is the 
velocity calculated from the particle momentum. As can be seen in Fig.~\ref{fig_random_coins}, 
there is a ``tail'' in the coincidence time spectrum 
at low $\beta$-$\beta_p$, which results mainly from pions undergoing nuclear 
interactions in the scintillators, aerogel or Cerenkov detector material. The 
produced slower hadrons are identified as pions, but generally have a larger 
time of flight. Furthermore there are pion events with $\beta=0$, meaning that 
no hits in the relevant scintillators were found when projecting the 
reconstructed track to the hodoscopes, which may also result from scattering 
of the pion.

The corrections for $\beta=0$ and the tail events were slightly different in 
Fpi-1 and Fpi-2. While the tail was neglected in Fpi-1, it was corrected for 
in Fpi-2. The latter approach includes the possibility of double-counting, 
when the tail particle was due to a pion that reacted in material, which was 
explicitly corrected for in the absorption calculation. Therefore, the 
absorption of pions and the various contributions in the 
$\beta$-$\beta_p$-spectrum were studied in more detail by calculating the 
number of pions reacting in various parts of the traversed material (also 
including elastic scattering), and estimating which fraction of these end-up 
where in the $\beta$-$\beta_p$ vs. coincidence-time spectrum. These studies 
were also performed for protons, where the absorption could be determined 
experimentally by comparing single (e,e) and coincident (e,e'p) events in the 
elastic peak in the measurements on the \heep\ reaction. The results indicated 
that the total transmission plus detection efficiencies for the used cuts 
differed by +1.8\% for Fpi-1 and -0.7\% for Fpi-2 from what had been used in 
the analysis. Since this is within the assumed uncertainty of the efficiency 
correction, and well within the overall uncertainty of the final separated 
cross sections, no additional correction was applied. 

\subsection{Backgrounds}
\label{sec:backgrounds}\nopagebreak[4]

\begin{figure}[!htb]
 \begin{center}
\includegraphics[width=2.5in]{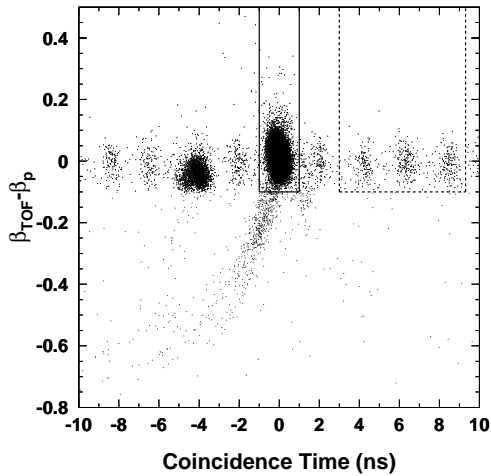}
 \caption{\label{fig_random_coins} \it (Color online) Coincidence time spectrum taken during Fpi-2, with the used real (solid) and random (dashed) coincidence time cuts. Real proton coincidences are clearly visible, but are rejected by the coincidence time cut. The tail is due to $\pi^+$ interactions in the detector elements, as explained
further in the text.}
 \end{center}
 \end{figure}

The coincidence timing structure between unrelated electrons and protons or pions from any 
two beam bursts is peaked every 2~ns, due to the accelerator timing structure. 
Real  and random $e$-$\pi$ coincidences were selected with cuts placed as shown in Fig.~\ref{fig_random_coins}. 
The random coincidence background during Fpi-1 was 2-5\%, depending on the kinematic setting, 
while it was always $<$ 1\% during Fpi-2.

The contribution of background events from the aluminum cell walls was estimated using dedicated 
runs with two ``dummy'' aluminium targets placed at the appropriate $z$-positions 
(see section~\ref{sec:target}). These data were analyzed in the same way as the cryotarget data 
and the yields were subtracted from the cryotarget yields,
taking into account the different thicknesses (about a factor of seven) of the target-cell walls 
and dummy target. The correction was small (2-4.5\%), while due to the high statistical 
accuracy of the dummy-target data, the contribution of the subtraction to the total uncertainty 
was negligible.

\subsection{Missing mass}
\label{sect:mm}

The reconstructed missing mass ($M_m$), see Fig.~\ref{fig:radtails},
provides an additional check on all momentum and angle calibrations. With the
calibrations and offsets discussed in sections~\ref{sec:optics} and~\ref{sec:offsets}
the values of the missing mass for the various kinematic cases were within
2 MeV of the neutron mass (with correction for radiative effects,
see section~\ref{sec:simc}). 
In the analysis a cut on the missing mass of $0.92 < M_m < 0.98$ GeV 
was used to ensure that no additional pions were produced.
The missing mass range was chosen in a region where the distribution is nearly flat (20~MeV above the missing mass peak), and resolution has a minimal effect on the yield,
and errors from insufficient simulation
of radiative processes at higher missing mass have not yet set in. Therefore, the result does not 
depend on the cut on the missing mass.

\section{Determination of the cross section}
\label{sec:xsecdet}\nopagebreak[4]

\subsection{Method}
\label{sec:method}\nopagebreak[4]

As described in section~\ref{subsec:cross_sections}, the (reduced) cross section can 
be written as a sum of four separate cross sections or structure functions,
which depend on $W$, \qsq\ and $t$,
\begin{eqnarray}
\label{eq:sepsig}
  2\pi \frac{d^2 \sigma}{dt d\phi} & = & \frac{d \sigma_T}{dt} + \epsilon  \frac{d \sigma_L}{dt} \\ \nonumber
                                   & + & \sqrt{2 \epsilon (1 + \epsilon)}  \frac{d \sigma_{LT}}{dt} cos \phi 
                                    +  \epsilon  \frac{d \sigma_{TT}}{dt} cos 2 \phi.
\end{eqnarray}

In order to be able to separate the different structure functions one has to
determine the cross section both at high and at low $\epsilon$ as a function
of the angle $\phi$ for fixed values of $W$, \qsq\ and $t$. Since the
$t$ dependence is important this should be done for various values of $t$
at every central \qsq\ setting.
Therefore, the data are binned in $t$ and $\phi$, thus integrating,
within the experimental acceptance,  over $W$ and \qsq, and also over $\theta_\pi$ 
(the latter is of relevance, since the interference structure
functions include a dependence on $\sin \theta_\pi$).
However, the average values of $W$, \qsq, and  $\theta_\pi$ generally
are not the same for different $\phi$ and for low and
high $\epsilon$. Moreover the average values of $W$, \qsq, $t$, and
$\theta_\pi$, only three of which are independent, may be inconsistent.

Both problems can be avoided by comparing the measured yields to
the results of a Monte-Carlo simulation for the actual experimental
setup (see the next section),
in which a realistic model of the cross section is implemented.
At the same time effects of finite experimental resolution, pion decay,
radiative effects, etc. can be taken into account.
When the model describes the dependence of the four structure functions on $W$,
\qsq, $t$, $\theta_\pi$ sufficiently well, i.e., when the ratio of 
experimental to simulated yields is close to unity within the statistical
uncertainty and does not depend on these variables anymore (except for a small linear dependence),
the cross section for any value of $\overline{W},\overline{Q^2}$ within the acceptance can be determined as 
\begin{equation}
\label{eq:ratio_to_sigma}
\left( \frac{d^2 \sigma}{dt d\phi}(t,\phi) \right)^{\mathrm{exp}}_{\overline{W},\overline{Q^2}}
=\frac{Y_{\mathrm{exp}}}{Y_{\mathrm{sim}}}
\left( \frac{d^2 \sigma}{dt d\phi}(t,\phi) \right)^{\mathrm{model}}_{\overline{W},\overline{Q^2}},
\end{equation}
where $Y$ is the yield over $W$ and $Q^2$, but common values of $\overline{W},\overline{Q^2}$ (if needed different for different
values of $t$) can be chosen for all values of $\phi$,  and for the high and low
$\epsilon$ data, so as to enable a separation of the structure functions.
In practice the data at both high and low $\epsilon$ were binned in 5 $t$-bins and 16 $\phi$-bins and the 
cross section was evaluated at the center of each bin. The overlined values in the expression above
were taken as the acceptance weighted average values for all $\phi$-bins (at both
high and low $\epsilon$) together, which
results in them being slightly different for the five $t$-bins.

\subsection{SIMC}
\label{sec:simc}\nopagebreak[4]

The Hall C Monte Carlo package SIMC has been used in the analysis of
several previous experiments, and is described in detail elsewhere
(see \eg~\cite{vol00, gaskell01}).
Only the key components (radiation, hadron decay, spectrometer optics, and multiple scattering) are presented here.
\begin{figure}[!htbp]
\centering
\includegraphics[width=3.5in]{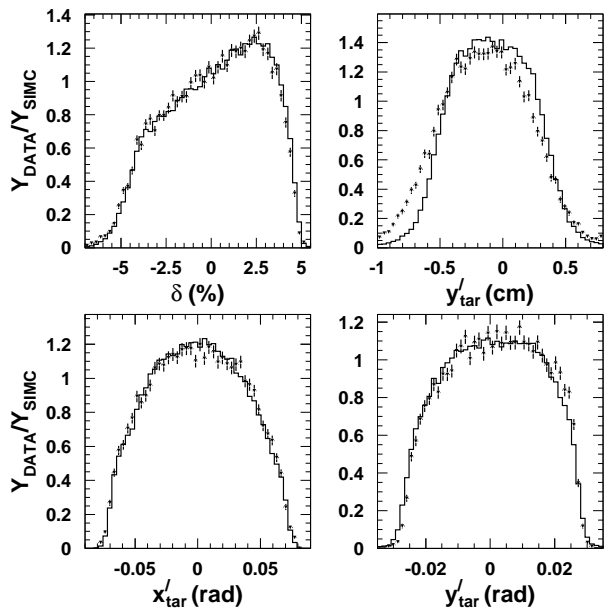}
\caption{\label{fig:distr_combined_recon} \it (Color online) Comparison of data (triangles) and SIMC (histogram) for HMS reconstructed quantities. The distributions were normalized to each other by one global scale factor.}
\end{figure}

For each event, the program generates the coordinates of the interaction vertex ($x,y,z$) 
and kinematic properties such as direction and momentum of the particles of interest. 
All angles are generated in the spectrometer coordinate system, where $z$ points in the direction of the beam, 
$x$ is vertical with x$>$0 pointing downwards and $y$ completes the right-handed coordinate system. 
The starting values for the generation are limited to a certain range, given as input.
When an event is kinematically  allowed, the event is radiated and the outgoing particles 
are followed on their way through the target, taking into account energy loss and multiple 
scattering.

After the event generation is complete, the events are sent to the single arm spectrometer 
modules, which simulate the magnetic optics inside the Hall C spectrometers using COSY~\cite{cosy} 
generated matrix elements\footnote{The COSY model consists of sets of 
``forward matrix elements'', which model the magnetic field in steps from one aperture 
to the next.}, and trace the particles through the magnetic fields, and rejecting events that
fall outside of several apertures along the spectrometer.

Simulated events that clear all apertures and cross the minimum number of detectors 
in the detector huts are considered to produce a valid trigger, and are reconstructed. The 
target quantities are reconstructed as described in section~\ref{sec:optics}, with realistic wire chamber resolutions and reconstruction matrix elements that are consistent with those used 
to trace the particles through the spectrometers.
Generally, the Monte Carlo simulation describes the data quite well, see Fig.~\ref{fig:distr_combined_recon}), 
except for small regions at the edges of the $ y'_\mathrm{tar}$ acceptance. A similar effect 
was observed in elastic scattering data. 
The $ y_\mathrm{tar}$ acceptance is not as well described. This quantity does, however, not
contribute to the calculation of any physics quantities, and was thus not further optimized.
Since only apertures are simulated, no inefficiencies 
are assigned in the event simulation. Finally each event is weighted by the relevant model 
cross section (see section~\ref{sec:modelxsec}) corrected for radiative processes, the overall 
luminosity, and a Jacobian taking into account the transformation between spectrometer 
and physics coordinates.

\begin{figure}[!htbp]
\centering
\includegraphics[width=3.5in]{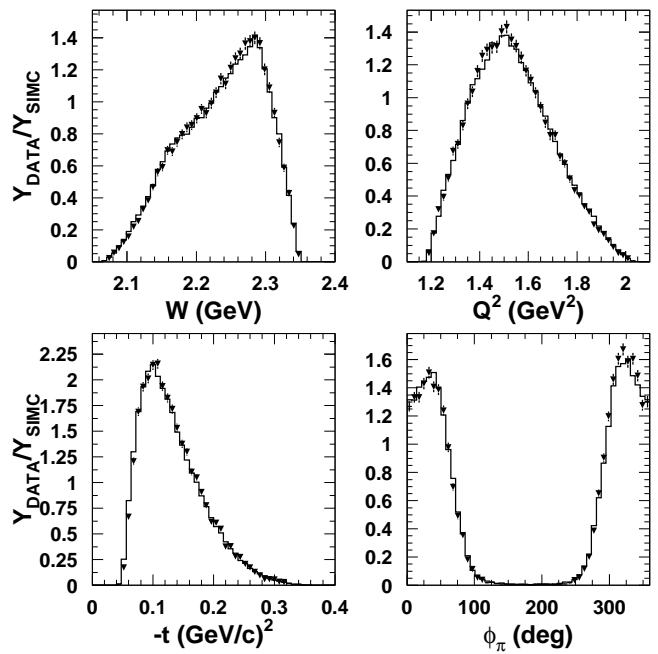}
\caption{\label{fig:distr_combined} \it  (Color online) Comparison of data (triangles) and SIMC (histogram) for the quantities $W$, $Q^2$, $-t$, and $\phi_\pi$. The distributions were normalized to each other by one global scale factor.}
\end{figure}

The reconstructed quantities are used in the comparison of the simulated and experimental 
distributions of various variables, 
an example of which is shown in Fig.~\ref{fig:distr_combined}. 
If the detector set-up is realistically simulated, the boundaries of measured and simulated 
distributions should match. Differences in magnitude can be attributed to differences between the 
actual cross section and the one used in the model.

Radiative effects describing the emission of real or virtual photons are an important 
part in the analysis of electron scattering data. The radiative corrections used in 
this analysis are based on the formalism of~\cite{Mo69}, and include both external and internal radiation.
The original formalism, derived for inclusive electron scattering, was extended for 
$(e,e^\prime p$) coincidence reactions in~\cite{Ent01}.

In calculating radiative processes for pion electroproduction, the target particle is 
a stationary proton and the final pion is taken to be an off-shell 
proton. The contribution from two-photon exchange diagrams is not included, but is 
expected to be very small~\cite{meln06}. 
The energy of the radiated photon is restricted to be much 
smaller than the energies of the inital and final state particles (soft photon 
approximation), and the radiation is taken to be in three discrete directions: along the 
direction of the incoming electron, of the scattered electron, and of the pion (extended 
peaking approximation).
\begin{figure}[!htb]
 \begin{center}
\includegraphics[width=3.5in]{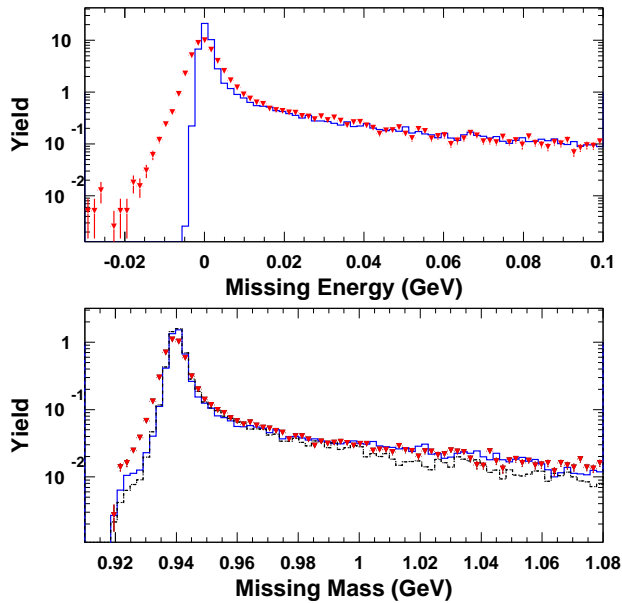}
 \caption{\label{fig:radtails} \it (Color online)
Top: Comparison between data (triangles) and SIMC (histogram) for the missing-energy 
distribution for one of the $^1$H($e,e^\prime p$) kinematics.
Bottom: Comparison between data (triangles) and SIMC (histogram) for the missing-mass distribution for 
a representative \heepi\ case. The solid histogram includes radiative effects and pions that pass through 
the HMS collimator. The latter events produce an additional contribution in the region $M_m$=1.025-1.07 GeV.
The dashed histogram represents simulated events without the effect of collimator punch-through.
}
 \end{center}
 \end{figure}

The method described above has been tested with $^1$H($e,e^\prime p$) data~\cite{mak94,Ent01}. 
An example for both $^1$H($e,e^\prime p$) and \heepi\ from the present experiment is 
shown in Fig.~\ref{fig:radtails}. The discrepancy at low missing energy for the $^1$H($e,e^\prime p$) case 
is due to an imperfect simulation of the resolution and peak shape in the tail. However, this does 
not influence the tail region. The simulated radiative tail gives a good description of the 
measured one. 
The global uncertainty is taken to be 2\%, with in case of pion production an additional 
uncertainty of 1\% to take into account the uncertainty associated with the extension of the 
formalism to pion electroproduction. 
The differential uncertainty in the L-T separation due to the radiative corrections was estimated 
by studying the integrated data/SIMC ratio as a function of the missing mass cut for different 
values of $\epsilon$. Although this ratio was found to vary up to 1.6\% when the cuts were 
applied, the dependence of the ratio on $\epsilon$ was relatively small.
Based on these studies a random uncertainty of 0.5\% between epsilon settings was assigned.

Charged pions decay into muons and (anti-)neutrinos with a branching fraction of 99.99\%. 
The fraction of pions decaying on their way from the target to the detection
system depends on their momentum and the pathlength, 
and was calculated to be up to 20\% for the lowest pion momenta. 
The possibility of pion decay in flight is included in SIMC, which accounts for events lost and for 
produced muons that still generate a valid trigger. A large fraction of the detected muons come 
from pion decay close to the target or pion decay in the field free region after the HMS magnetic 
elements and inside the spectrometer hut. 
About 4\% of the events detected in the spectrometer result from pions that have decayed in flight.
The overall uncertainty due to the simulation of pion decay was taken to be 1\%.
Since the pion momentum distributions are very similar between high and low epsilon settings, the 
random uncertainty between $\epsilon$ settings is very small (about 0.03\%), mainly accounting for 
muons coming from pions normally outside the acceptance.

\subsubsection {Checks with \heep\ }
\label{sec:simc_heep}\nopagebreak[4]

In addition to providing information on experimental offsets (see 
section~\ref{sec:offsets}), the elastic \heep\ reaction also serves to 
check the accuracy of the phase space model in SIMC, and, since
the elastic cross section is well known, it can be used to 
study the accuracy of the calculated yields.

In  Fpi-1 (Fpi-2), data for the elastic \heep\ reaction were taken in
five (four) different kinematic settings, all of which were modeled in SIMC.
The experimental and simulated missing energy distributions for one of the
settings were already shown in Fig.~\ref{fig:radtails}.
Also other simulated distributions were in good agreement with the
experimental data in all cases except for
the kinematic setting in which the SOS is at an angle of 56$\deg$.
It was found that the model for the SOS in SIMC does not describe correctly
the acceptance for part of the events when $|y_{tar}|$ becomes large
(see section~\ref{sec:acceptance}).
When that particular region of the phase space was removed 
from the analysis, the agreement was similar as for the other kinematics. 

\begin{figure}[!htbp]
\centering
\includegraphics[width=3.5in]{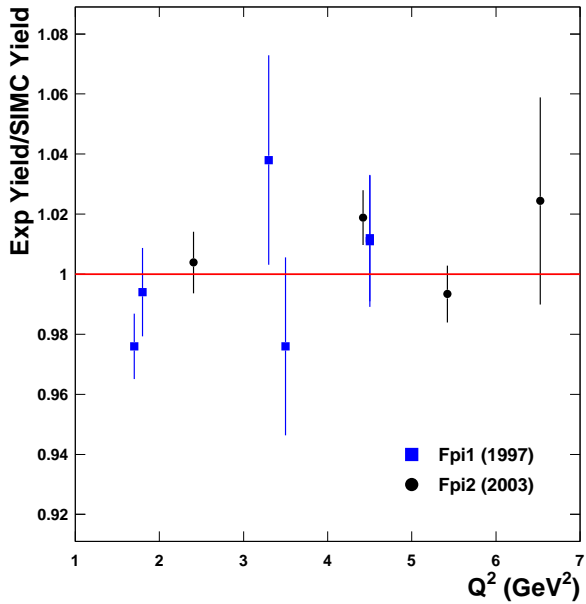}
\caption{\label{fig:HeepYields} \it (Color online) 
	The yield ratios from elastic data and SIMC for Fpi-1 and Fpi-2. The error bars include 
        statistical uncertainties only. The systematic uncertainty is about 2\%.}
\end{figure}

The total measured and simulated yields are compared in Fig.~\ref{fig:HeepYields}. 
The elastic cross sections used in the simulation were taken from the fit 
to the world data of ~\cite{bos95} for Fpi-1. For Fpi-2, the improved fit
from ~\cite{arr04} was also considered. In the region of interest, differences 
between the two are less than 2.0\%.
Over the whole $Q^2$ range between 1.5 and 5.4~\gevsq\, the ratio scatters
around unity with $\sigma$=2.0\%, consistent with the uncertainty of the
individual points.
In addition, one should take into account the uncertainty of the world's
data, which is of comparable magnitude.

These results demonstrate that the efficiencies and dead times used to
calculate the experimental yields are well understood, and that the Monte Carlo
program simulates the experimental conditions and acceptances very well.

\subsubsection{Detector acceptances in SIMC}
\label{sec:acceptance}\nopagebreak[4]

In the \heep\ reaction the outgoing electron and proton are
strongly correlated, so that only a subset of the phase space is populated. 
The full SOS acceptance was studied by measuring deep-inelastic electron 
scattering from deuterium. A detailed comparison of the boundaries of the 
acceptance of the experimental and the simulated data in the four 
target variables  $\delta, y_\mathrm{tar}, x'_\mathrm{tar}$, and 
$y'_\mathrm{tar}$ (see Subsect.~\ref{sec:optics})
revealed that in the region: \\
        $y'_\mathrm{tar}>(-125.0+4.25 \hspace{1mm} \delta+64.0 
	\hspace{1mm}
        y_\mathrm{tar}-1.7 \hspace{1mm} \delta \hspace{1mm} y_\mathrm{tar}$), 
	and \\
        $y'_\mathrm{tar}< (\hspace{2.7mm} 125.0-4.25 \hspace{1mm} 
	\delta+64.0
        \hspace{1mm} y_\mathrm{tar}-1.7 \hspace{1mm} \delta \hspace{1mm}
        y_\mathrm{tar}$) \\
with $y'_\mathrm{tar}$ in mrad, $y_\mathrm{tar}$ in cm, $\delta$ in \%
the boundaries did not match, with SIMC losing events that were present in the
data. Therefore, these parts of the acceptance were excluded from the analysis.

The model for the HMS acceptance does not present a comparable challenge.
As the HMS is placed at very forward angles in all kinematics, the 
$y_\mathrm{tar}$ acceptance is flat in the (limited) region of interest. 
The acceptances in $y'_\mathrm{tar}$ and $\delta$ used in the
analysis, 
are within the previously determined safe boundaries.
The phase space (boundaries) for coincident HMS and SOS events was checked
with data from the pion electroproduction reaction by comparing distributions
for quantities such as HMS and SOS reconstructed target variables, $W$, $Q^2$,
$t$, and missing energy and momenta, see Figs.~\ref{fig:distr_combined_recon} 
\ref{fig:distr_combined}, and \ref{fig:HeepYields}.

The uncertainties due to spectrometer acceptance was tested by varying the cuts on the 
quantities ($\delta$, $x^\prime_{tar}$, $y^\prime_{tar}$) in each spectrometer. The 
experimental cross section was then extracted for spectrometer cut variations of $\pm$ 10\% 
and compared to the one with nominal cuts. In general, the variation of the cross 
section is small ($<$ 0.5\%).

\subsubsection{The model cross section}
\label{sec:modelxsec}\nopagebreak[4]

\begin{figure}[!htb]
 \begin{center}
\includegraphics[width=3.5in]{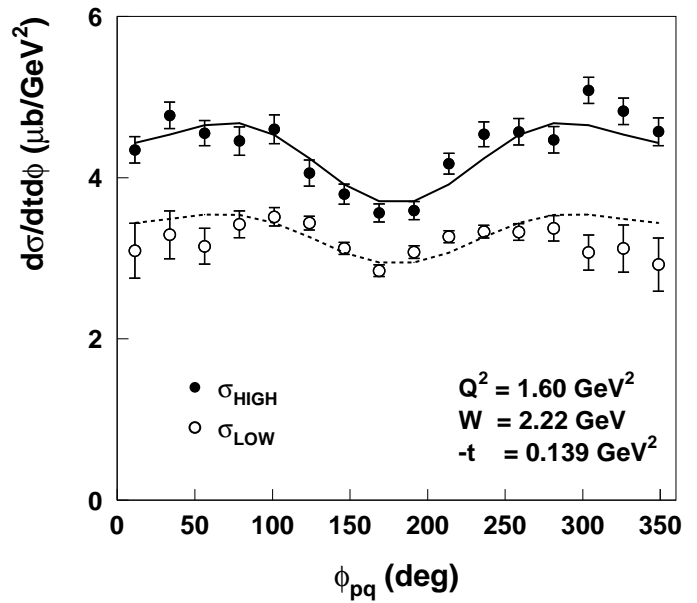}
 \caption{\label{fig_unpol_xsec} \it (Color online) Representative plot of the experimental cross sections,
$\frac{d^2 \sigma}{dt d\phi}$ as a function of the azimuthal angle $\phi_\pi$ at $Q^2$=1.60
(GeV$^2$) for high and low $\epsilon$. The curves shown represent the fit of the measured
values of the cross section to equation~\protect{~\ref{eq:sepsig}}.}
 \end{center}
 \end{figure}

The model cross section and the final separated structure functions were determined 
in the same (iterative) procedure. The model cross section was taken as the product 
of a global function describing the $W$-dependence times (a sum of) \qsq\ and $t$ 
dependent functions for the different structure functions.
For the LT and TT parts, their leading order dependence on $\sin$($\theta^*$) was 
taken into account~\cite{sintheta}. The $W$-dependence was taken as $(W^2-M^2_p)^{-2}$, based on analyses 
of experimental data from~\cite{bra77,beb78}. For the parts depending on $Q^2$ and $t$, 
phenomenological forms were used and the parameters were fitted. For all five $t$-bins 
at every (central) \qsq\ setting, $\phi$-dependent cross sections were determined both 
at high and low $\epsilon$ for chosen values of $\overline{W},\overline{Q}^2$ (and 
corresponding values of $\theta_{\pi}$ and $\epsilon$) according to
\begin{equation}
   \label{eq:ratio_to_sigphi}
   \sigma_{\mathrm{exp}}(\overline{W},\overline{Q}^2,t,\phi;\overline{\theta},\overline{\epsilon}) =
   \frac{\langle Y_{\mathrm{exp}}\rangle}{\langle Y_{\mathrm{sim}}\rangle} \,\,
   \sigma_{\mathrm{MC}}(\overline{W},\overline{Q}^2,t,\phi;\overline{\theta},\overline{\epsilon}).
\end{equation}
The fitting procedure was iterated until $\sigma_{exp}$ changed by less than a prescribed amount (typically 1\%).
A representative example of the experimental cross section and the fit as a function of $\phi_\pi$
 is shown in figure ~\ref{fig_unpol_xsec}.
The cosine structure from the interference terms is clearly visible.

This procedure was carried out independently for Fpi-1 and Fpi-2 in order to have optimal descriptions in the two different kinematic ranges covered~\footnote{These parameterizations are for a nominal value of $W$=1.95 GeV}. The final cross section parameterization for Fpi-1 (the cross sections have units of $\mu b/\mathrm{GeV}^2$, and the units of $Q^2$, $t$, and $m_{\pi}^2$ are GeV$^2$) is:
\begin{eqnarray}
\frac{d\sigl}{dt} & = & 36.51 \ e^{(26.10-7.75 Q^2 ) (t+0.02)}\\ \nonumber
\label{eq:piongent}
\frac{d\sigt}{dt} & = & \frac{0.74}{Q^2} + \frac{1.25}{Q^4} + 0.57 \ \frac {|t|}{(|t|+ m_{\pi}^2)^2} \\ \nonumber
\label{eq:piongenlt}
\frac{d\siglt}{dt} & = & \left (e^{(4.69+\frac{24.55}{\sqrt{Q^2}} \ t)}+1.47-\frac{7.89}{Q^4} \right ) \ \sin \theta^*  \\ \nonumber
\label{eq:piongentt}
\frac{d\sigtt}{dt} & = & \left(\frac{3.44}{Q^2}-\frac{7.57}{Q^4} \right) \cdot \frac {|t|}{(|t|+ m_{\pi}^2)^2} \ \sin^2{\theta^\ast}.
\end{eqnarray}
This parameterization is valid in the range \qsq\ between 0.4 and 1.8 \gevsq. 

The Fpi-2 parameterization, valid between $Q^2= 1.4$ and 2.7 \gevsq, is:
\begin{eqnarray}
\label{eqn_parm04}
  \frac{d\sigl}{dt}  &=&  \frac{350 \ Q^2}{(1+1.77 Q^2+0.05 Q^4)^2} \ e^{(16-7.5 \ln Q^2) t} \\ \nonumber
  \frac{d\sigt}{dt}  &=&  \frac{4.5}{Q^2}+\frac{2.0}{Q^4} \\ \nonumber
  \frac{d\siglt}{dt} & = & \left (e^{(0.79+\frac{3.4}{\sqrt{Q^2}} \ t)}+1.1-\frac{3.6}{Q^4} \right ) \ \sin \theta^*  \\ \nonumber
  \frac{d\sigtt}{dt} &=&  -\frac{5.0}{Q^4} \ \frac {|t|}{(|t|+ m_{\pi}^2)^2} \ \sin^2 \theta^* 
\end{eqnarray}

Since the extracted separated cross sections depend in principle on the cross section model, there is a ``model'' systematic uncertainty. This uncertainty was studied by extracting \sigl\ and \sigt\ with different cross section models. Since the longitudinal and transverse cross sections in the model reproduce the experimental values to within 10\%, these two terms were independently increased and decreased by 10\% in the model. With these changes, the extracted \sigl\ and \sigt\ varied by less than 0.5\%. For evaluating the model uncertainty due to the interference terms \siglt\ and \sigtt\, these terms were independently increased or decreased by their respective uncertainties, obtained when fitting the four structure functions, and L/T separations were done with the modified models. The contribution to the uncertainty of \sigl\ and \sigt\ of these two terms is between 1\% and 8\% and depends strongly on $t$. The latter value (at the largest values of $-t$) is comparable to the contribution of uncorrelated uncertainties to \sigl\ and \sigt.

\subsection{Estimate of uncertainties}
\label{sec:errorpropagation}\nopagebreak[4]

The statistical uncertainties in the unseparated cross sections are determined by the uncertainties 
in $Y_{\mathrm{exp}}$ and $Y_{\mathrm{sim}}$ in Eq.~(\protect\ref{eq:ratio_to_sigma}). The statistical 
uncertainty in $R=Y_{\mathrm{exp}}/Y_{\mathrm{sim}}$ is dominated by the uncertainty in the number of 
measured real events, and ranges from 1\% to 3\%, depending on the values of \qsq\ and $t$.

The systematic uncertainties can be subdivided into correlated and uncorrelated contributions. The 
correlated uncertainties, i.e., those that are the same for both epsilon points, such as target thickness
corrections, are attributed directly to the separated cross sections. Uncorrelated uncertainties are 
attributed to the unseparated cross sections, with the result that in the separation of \sigl\ and \sigt\ 
they are inflated, just as the statistical uncertainties, by the factor $1/\Delta\epsilon$ (for $\sigma_L$), which is 
about three. They can be further subdivided into uncertainties that differ in size between $\epsilon$ points, 
but may influence the $t$-dependence at a fixed value of $\epsilon$ in a correlated way.

All systematic uncertainties for Fpi-2, with their subdivison, are listed in table~\ref{tab:syst_unc}.
They have been added quadratically to obtain the total systematic uncertainty.
For Fpi-1 the values are similar and only the total systematic uncertainties
for the different categories are given.
The ``instrumental'' and model uncertainties have been already discussed in previous (sub)sections.
The uncertainties in the acceptance are based on extensive single-arm elastic and deep-inelastic measurements, 
both from the present experiment and from~\cite{tva04,chr04}, and \heep\ data, plus how well the sieve-slit 
is reproduced by the used optical matrix elements.
The influence of the uncertainties in the offsets in the kinematical variables
such as beam energy, momenta and angles, were determined by changing the latter
by their uncertainty and evaluating the resultant changes in the separated cross sections.

The largest fully correlated systematic uncertainties are the ones due to the
radiative corrections, pion absorption, and pion decay,
resulting in a total correlated uncertainty of 3-4\%. 
The fully uncorrelated systematic uncertainty is dominated by acceptance,
resulting in a total uncorrelated uncertainty of 0.7 to 1.2\%.
The largest contributions to the ``t-correlated'' uncertainty are acceptance,
model dependence, and kinematic offsets, resulting in a total $\epsilon$ uncorrelated, $t$ correlated 
uncertainty of 1.7 to 2.0\%.
As mentioned, these $\epsilon$ uncorrelated uncertainties are multiplied by
about a factor of three when performing the L/T separation. As a result,
they are the dominating systematic uncertainty for, e.g., \sigl.
 
\begin{tiny}
 \begin{table}[!ht]
  \renewcommand{\arraystretch}{1.2}
  \centering
  \begin{tabular}{||l|c|c|c|c||}
  \hline  
 Correction       & Uncorr.           & $\epsilon$ uncorr.  & Corr.     & Section   \\
                  & (pt-to-pt)        & $t$ corr.           &  (scale)  &           \\
                  &  (\%)             & (\%)                &     (\%)  &           \\
 \hline
 \hline  
 Acceptance        &    1.0 (0.6)      &     0.6          &  1.0        & \ref{sec:acceptance}   \\
 Model Dep         &    0.2            &     1.1-1.3      &  0.5        & \ref{sec:modelxsec}    \\
 d$\theta_{e}$     &    0.1            &     0.7-1.1      &             & \ref{sec:offsets}      \\
 d$E_{beam}$       &    0.1            &     0.2-0.3      &             & \ref{sec:offsets}      \\
 d$P_{e}$          &    0.1            &     0.1-0.3      &             & \ref{sec:offsets}      \\ 
 d$\theta_{\pi}$   &    0.1            &     0.2-0.3      &             & \ref{sec:offsets}      \\
 Radiative corr    &                   &      0.4          &  2.0       & \ref{sec:simc}         \\
 Pion absorption   &                   &      0.1         &  2.0        & \ref{sec:absorption}   \\
 Pion decay        &    0.03           &                  &  1.0        & \ref{sec:simc}         \\
 HMS Tracking      &                   &      0.4         &  1.0        & \ref{sec:tracking_eff} \\
 SOS Tracking      &                   &      0.1         &  0.5        & \ref{sec:tracking_eff} \\
 Charge            &                   &      0.3         &  0.4        & \ref{sec:accelerator}  \\
 Target Thickness  &                   &      0.2         &  0.9        & \ref{sec:targetdens}   \\
 CPU dead time     &                   &      0.2         &             & \ref{subsec:deadtimes}   \\
 HMS Trigger       &                   &      0.1         &             & \ref{subsect:triggereff} \\
 SOS Trigger       &                   &      0.1         &             & \ref{subsect:triggereff} \\
 Ele DT            &                   &      0.3         &             & \ref{subsec:deadtimes}   \\
 Coincidence block. &                  &      0.1         &             & \ref{sect:coinblock}     \\
 Particle ID       &                   &      0.2         &             & \ref{sec:pid}            \\ 
\hline  
 Total (Fpi-2)            &    1.2 (0.9) &     1.8-1.9     &  3.5        & \ref{sec:errorpropagation}             \\
\hline  
 Total (Fpi-1)            &    0.7       &     1.7-2.0     &  2.8        & \ref{sec:errorpropagation}             \\
\hline  
  \end{tabular}
 \caption{\label{tab:syst_unc} \it Summary of systematic uncertainties for Fpi-2.
Where two values are given, they are for the two \qsq\ points. 
When a range is given, it corresponds to the range in $t$-values.
The last column gives the sections where the various items are discussed.
For Fpi-1 only the total uncertainties are listed as the individual contributions are similar to those from Fpi-2.
}
 \end{table}
\end{tiny}

\section{Cross section results}
\label{sec:results}\nopagebreak[4]

\begin{figure}[!htbp]
\centering
\includegraphics[width=3.5in]{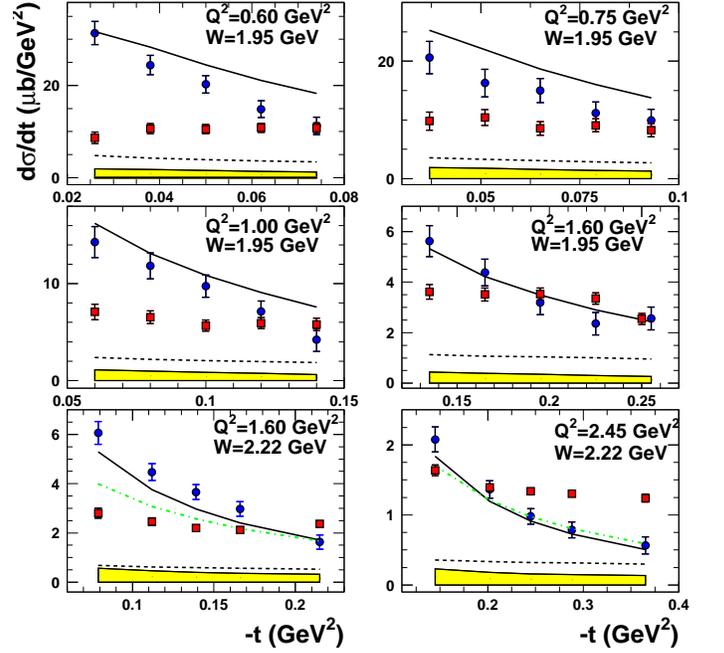}
\caption{\label{fig:xsec1H} \it The separated cross sections, \sigL\ (circles) and
  \sigT\ (squares) at central values of $Q^2$=0.60, 0.75, 1.00, 1.60 GeV$^2$
  ($W$=1.95~GeV), and $Q^2$=1.60, 2.45 GeV$^2$ ($W$=2.22~GeV). The values
  of $\overline{W}$ and $\overline{Q^2}$ are different for each
  $-t$-bin. The error bars for \sigL
  indicate the statistical and uncorrelated systematic uncertainties in both
  $\epsilon$ and $-t$ combined in quadrature. The error band denotes the
  correlated part of the systematic uncertainty by which all data points move
  collectively for \sigL. The error bars for \sigT represent the total uncertainty. The curves denote Regge calculations 
  (VGL,~\cite{van97}) for \sigL (solid
  line) and \sigT (dashed line) for $\Lambda_{\pi}^2$=0.462 GeV$^2$ and
  $\Lambda_{\rho}^2$=1.5 GeV$^2$. Also shown is a calculation for \sigL
  (dashed-dotted line) using a GPD model~\cite{Vdh98} including power corrections.}
\end{figure}

\begin{figure}[!htbp]
\centering
\includegraphics[width=3.5in]{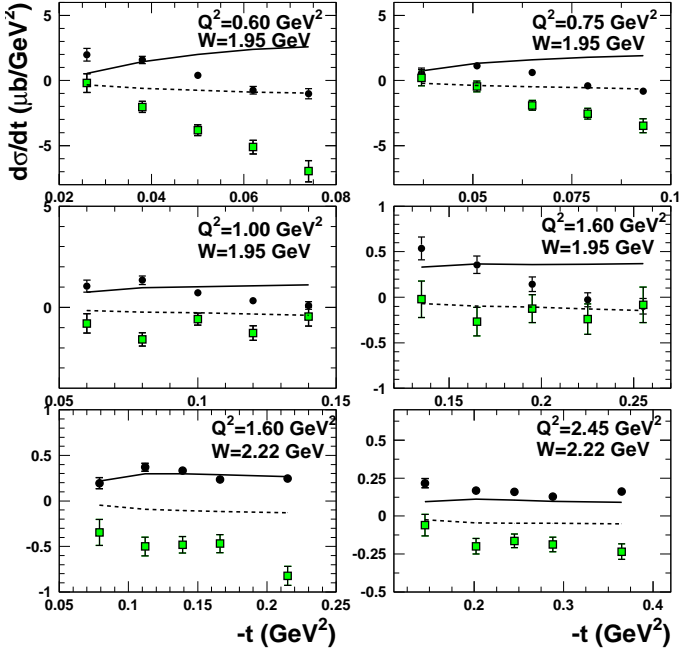}
\caption{\label{fig:xsec2H} \it  The interference terms, \sigLT\ (circles) and
  \sigTT\ (squares) at central values of $Q^2$=0.60, 0.75, 1.00, 1.60 GeV$^2$
  ($W$=1.95~GeV), and $Q^2$=1.60, 2.45 GeV$^2$ ($W$=2.22~GeV). 
  The curves denote Regge calculations 
  (VGL,~\cite{van97}) for \sigLT\ (solid line) and \sigTT\ (dashed
  line) with $\Lambda_{\pi}^2$=0.462 GeV$^2$ and $\Lambda_{\rho}^2$=1.5 GeV$^2$.}  
\end{figure}

The separated cross sections are listed in Table~\ref{tab:xsecH} and 
shown in Figs.~\ref{fig:xsec1H} (\sigL, \sigT) and \ref{fig:xsec2H}
(\sigLT, \sigTT). 
In the following subsections, the global dependences of \sigL\ and \sigT\ will
be reviewed, and the data compared to model calculations for the
\heepi\ reaction.

 \begin{small}
  \begin{table*}
 \begin{center}  
  \begin{tabular}{||c|c|c|c|c|c|c||}
  \hline  
 $\overline{Q^2}$ & $\overline{W}$ & $-t$ & \sigl      & \sigt         & \siglt             & \sigtt  \\
 (GeV$^2$)  & (GeV) & (GeV$^2$)  & ($\mu$b/GeV$^2$)          & ($\mu$b/GeV$^2$)      & ($\mu$b/GeV$^2$)          & ($\mu$b/GeV$^2$)\\
 \hline
\multicolumn{7}{|c|}{$Q^2=0.60$ \gevsq\  $W=1.95$ GeV} \\
 \hline  
 0.526        & 1.983 & 0.026      & 31.360 $\pm$ 1.602, 1.927 & 8.672  $\pm$ 1.241  & 1.982 $\pm$ 0.491  & -0.187  $\pm$ 0.71     \\
 0.576        & 1.956 & 0.038      & 24.410 $\pm$ 1.119, 1.774 & 10.660 $\pm$ 1.081  & 1.581 $\pm$ 0.288  & -2.034 $\pm$ 0.427     \\
 0.612        & 1.942 & 0.050      & 20.240 $\pm$ 1.044, 1.583 & 10.520 $\pm$ 1.000  & 0.409 $\pm$ 0.255  & -3.811 $\pm$ 0.406     \\
 0.631        & 1.934 & 0.062      & 14.870 $\pm$ 1.155, 1.366 & 10.820 $\pm$ 0.992  & -0.745 $\pm$ 0.302 & -5.117 $\pm$ 0.524     \\
 0.646        & 1.929 & 0.074      & 11.230 $\pm$ 1.469, 1.210 & 10.770 $\pm$ 1.097  & -1.020 $\pm$ 0.390 & -6.966 $\pm$ 0.816     \\
 \hline  
\multicolumn{7}{|c|}{$Q^2=0.75$ \gevsq\  $W=1.95$ GeV} \\
 \hline  
 0.660        & 1.992 & 0.037      & 20.600 $\pm$ 1.976, 1.895 & 9.812  $\pm$ 1.532  &  0.565 $\pm$ 0.393  & 0.208  $\pm$ 0.623    \\
 0.707        & 1.961 & 0.051      & 16.280 $\pm$ 1.509, 1.788 & 10.440 $\pm$ 1.344  &  1.135 $\pm$ 0.268  & -0.454 $\pm$ 0.420    \\
 0.753        & 1.943 & 0.065      & 14.990 $\pm$ 1.270, 1.573 & 8.580  $\pm$ 1.150  &  0.618 $\pm$ 0.206  & -1.910  $\pm$ 0.378   \\
 0.781        & 1.930 & 0.079      & 11.170 $\pm$ 1.214, 1.416 & 9.084  $\pm$ 1.091  &  -0.409 $\pm$ 0.197 & -2.547  $\pm$ 0.419   \\
 0.794        & 1.926 & 0.093      &  9.949 $\pm$ 1.376, 1.277 & 8.267  $\pm$ 1.110  &  -0.827 $\pm$ 0.220 & -3.474  $\pm$ 0.534   \\
 \hline  
\multicolumn{7}{|c|}{$Q^2=1.00$ \gevsq\  $W=1.95$ GeV} \\
 \hline  
 0.877        & 1.999 & 0.060      & 14.280 $\pm$ 1.157, 1.103 & 7.084  $\pm$ 0.791  &  1.049 $\pm$ 0.294 & -0.794  $\pm$ 0.474    \\
 0.945        & 1.970 & 0.080      & 11.840 $\pm$ 0.887, 0.978 & 6.526  $\pm$ 0.657  &  1.339 $\pm$ 0.205 & -1.584  $\pm$ 0.329    \\
 1.010        & 1.943 & 0.100      &  9.732 $\pm$ 0.773, 0.837 & 5.656  $\pm$ 0.572  &  0.719 $\pm$ 0.164 & -0.582  $\pm$ 0.302    \\
 1.050        & 1.926 & 0.120      &  7.116 $\pm$ 0.789, 0.747 & 5.926  $\pm$ 0.570  &  0.331 $\pm$ 0.158 & -1.277  $\pm$ 0.360    \\
 1.067        & 1.921 & 0.140      &  4.207 $\pm$ 1.012, 0.612 & 5.802  $\pm$ 0.656  &  0.087 $\pm$ 0.187 & -0.458  $\pm$ 0.471    \\
 \hline  
\multicolumn{7}{|c|}{$Q^2=1.60$ \gevsq\  $W=1.95$ GeV} \\
 \hline  
 1.455        & 2.001 & 0.135      &  5.618 $\pm$ 0.431, 0.442 & 3.613  $\pm$ 0.294  &  0.537 $\pm$ 0.125 & -0.022  $\pm$ 0.200    \\
 1.532        & 1.975 & 0.165      &  4.378 $\pm$ 0.356, 0.390 & 3.507  $\pm$ 0.257  &  0.356 $\pm$ 0.095 & -0.268  $\pm$ 0.156    \\
 1.610        & 1.944 & 0.195      &  3.191 $\pm$ 0.322, 0.351 & 3.528  $\pm$ 0.241  &  0.143 $\pm$ 0.081 & -0.126  $\pm$ 0.153    \\
 1.664        & 1.924 & 0.225      &  2.357 $\pm$ 0.313, 0.310 & 3.354  $\pm$ 0.228  &  -0.028 $\pm$ 0.076 & -0.241  $\pm$ 0.167   \\
 1.702        & 1.911 & 0.255      &  2.563 $\pm$ 0.356, 0.268 & 2.542  $\pm$ 0.227  &  -0.100 $\pm$ 0.085 & -0.083  $\pm$ 0.196   \\
 \hline  
\multicolumn{7}{|c|}{$Q^2=1.60$ \gevsq\  $W=2.22$ GeV} \\
 \hline  
 1.416        & 2.274  & 0.079      & 6.060 $\pm$ 0.464, 0.564 & 2.802   $\pm$ 0.27   &  0.195 $\pm$ 0.073 & -0.346 $\pm$ 0.177     \\
 1.513        & 2.242 & 0.112      & 4.470 $\pm$ 0.342, 0.457 & 2.459   $\pm$ 0.21   &  0.370 $\pm$ 0.081 & -0.500 $\pm$ 0.169     \\
 1.593        & 2.213 & 0.139      & 3.661 $\pm$ 0.303, 0.397 & 2.198   $\pm$ 0.19   &  0.334 $\pm$ 0.089 & -0.481 $\pm$ 0.139     \\
 1.667        & 2.187 & 0.166      & 2.975 $\pm$ 0.294, 0.358 & 2.124   $\pm$ 0.18   &  0.235 $\pm$ 0.081 & -0.469 $\pm$ 0.139     \\
 1.763        & 2.153 & 0.215      & 1.630 $\pm$ 0.292, 0.315 & 2.369   $\pm$ 0.19   &  0.247 $\pm$ 0.087 & -0.823 $\pm$ 0.300     \\
 \hline  
\multicolumn{7}{|c|}{$Q^2=2.45$ \gevsq\  $W=2.22$ GeV} \\
 \hline  
 2.215        & 2.308 & 0.145      & 2.078 $\pm$ 0.180, 0.229 & 1.635   $\pm$ 0.11   &  0.217 $\pm$ 0.034 & -0.060 $\pm$ 0.163     \\
 2.279        & 2.264 & 0.202      & 1.365 $\pm$ 0.125, 0.179 & 1.395   $\pm$ 0.08   &  0.168 $\pm$ 0.025 & -0.199 $\pm$ 0.066     \\
 2.411        & 2.223 & 0.245      & 0.980 $\pm$ 0.110, 0.159 & 1.337   $\pm$ 0.08   &  0.159 $\pm$ 0.023 & -0.163 $\pm$ 0.045     \\
 2.539        & 2.181 & 0.288      & 0.786 $\pm$ 0.114, 0.150 & 1.304   $\pm$ 0.08   &  0.128 $\pm$ 0.018 & -0.187 $\pm$ 0.120     \\
 2.703        & 2.127 & 0.365      & 0.564 $\pm$ 0.123, 0.137 & 1.240   $\pm$ 0.08   &  0.161 $\pm$ 0.020 & -0.234 $\pm$ 0.109     \\
\hline  
  \end{tabular}
  \end{center}
 \caption{\label{tab:xsecH} \it Separated cross sections
 \sigL, \sigT, \sigLT, and \sigTT\ for the \heepi\ reaction for Fpi-1 and Fpi-2.
 The two uncertainties given for \sigL\ are the combination of statistical and $t$-uncorrelated
 systematic uncertainties, and the combination of the $\epsilon$-correlated (scale) 
 and $\epsilon$-uncorrelated, $t$-correlated uncertainties.
This distinction is relevant when extracting values of \fpi\ from the measured
values of \sigl\ (see~\cite{hub08}).
The uncertainties for \sigT, \sigLT, and \sigTT\ include all uncertainties.}
 \end{table*}
\end{small}

\subsection{Global dependences of the separated cross sections}

At all values of $Q^2$, the longitudinal cross section \sigL\ shows the
characteristic fall-off with $-t$ due to the pion pole. Its magnitude
(at constant $W$)  drops with increasing $Q^2$, mainly because the value
of $-t_{min}$ increases with \qsq.
The transverse cross section \sigT\ is largely flat with $-t$, while its
magnitude drops with increasing \qsq.
The interference term \sigLT\ is rather small, while the value of \sigTT, which
clearly shows the behavior of going to 0 at $t_{min}$, drops rapidly with
increasing \qsq.

With the availability of our precision separated cross sections over an
extended kinematic range, it is interesting to look into the global dependences
of the longitudinal and transverse cross sections upon $W$, $Q^2$ and $t$.
A similar study was done in~\cite{bra76} with the more limited data then
available.  For the purpose of this study, both our cross sections and those
of~\cite{ack78,bra76} were used.  For \sigt, the photoproduction
data of~\cite{boy68} were also used.

The $W$ dependences of the earlier \sigl\ and \sigt\ data were observed~\cite{bra76}
to follow $(W^2 - M^2)^{-2}$, where $M$ is the nucleon mass.  Our $Q^2$=1.60 \gevsq\ data
at $W$=1.95, 2.22 GeV are consistent with this within about 10\%. 

\begin{figure}
\begin{center}
\includegraphics[width=3.5in]{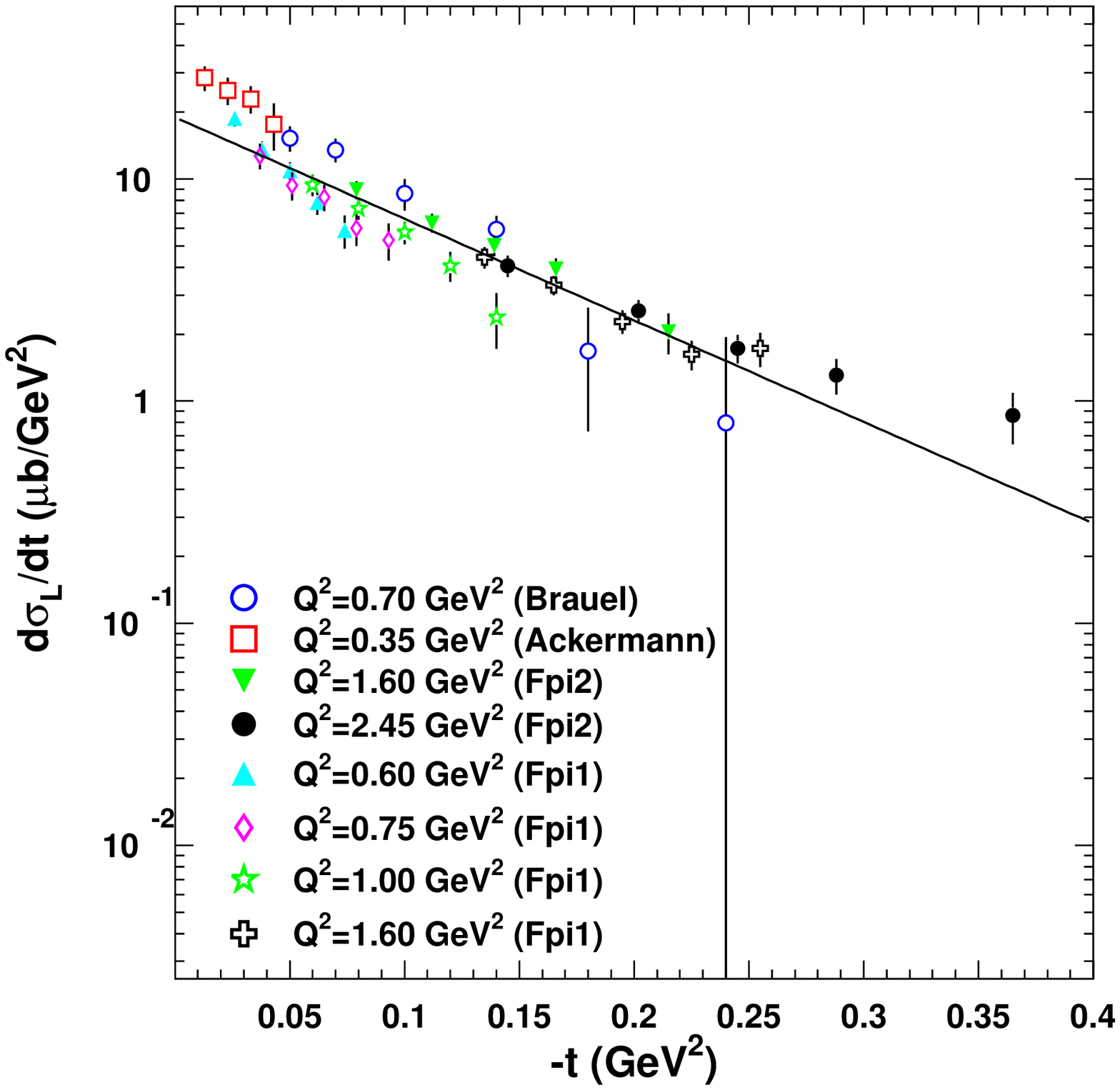}
\caption{\label{fig-sigl-tdep-all} \it (Color online) The $t$ dependence of the longitudinal
  $\pi^+$ cross section. The data from ~\cite{ack78,bra76,tad06,horn06} are scaled
  in $W$ and $Q^2$ (see the text) to common values of $W$=2.19 GeV and $Q^2=$0.7 \gevsq.}
\end{center} 
\end{figure}

Because \sigl\ is dominated by the pion-pole contribution, its
$Q^2$-dependence is largely given by $Q^2 F_{\pi}^2(Q^2)$. 
Fig.~\ref{fig-sigl-tdep-all} shows the results for \sigL, where
all cross sections have been scaled to $W$=2.19 GeV according
to $(W^2 - M^2)^{-2}$, and to $Q^2$=0.70~GeV$^2$ using the factor $Q^2 F_{\pi}^2(Q^2)$,
where $F_{\pi}$ was assumed to follow the monopole form $(1+\frac{Q^2}{m^2_{\rho}})^{-1}$.
Although overall the \sigl\ data follow an almost exponential $t$ dependence,
upon close inspection it is observed that at constant \qsq\ the data
deviate from that curve, i.e., the \qsq\ and $t$-dependences do not factorize
completely, and at both high and low $-t$ deviations from a pure exponential are
observed.
Fitting  the data with an exponential $B e^{-b|t|}$ results in a slope
parameter $b$=10.5 $\pm$ 1.8
GeV$^{-2}$, and a normalization factor $B$=19.0 $\pm$ 2.0.  Such a form
describes all \sigL\ data within about 50\%. 

\begin{figure}[!htb]
\begin{center}
\includegraphics[width=3.5in]{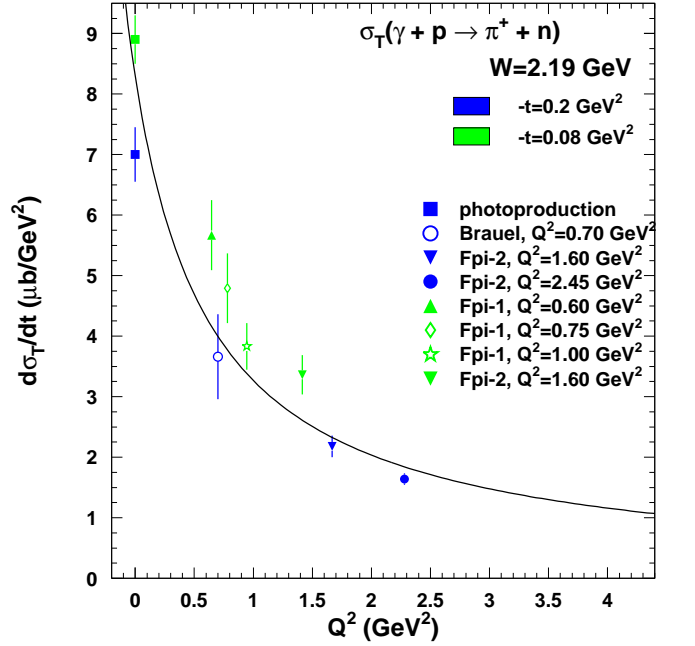}
\caption{\label{fig-sigt-q2dep-all} \it (Color online) The $Q^2$ dependence of the transverse
  $\pi^+$ cross section at $-t$=0.08 and 0.2 GeV$^2$. The cross sections are scaled to
  $W$=2.19 GeV. The photoproduction point is from~\cite{boy68}.
  The curve indicates a parameterization for \sigT\ of the form $\frac{C}{1+DQ^2}$.}
\end{center}
\end{figure}

No simple prediction exists for the \qsq-dependence of \sigT.  
Fig.~\ref{fig-sigt-q2dep-all} shows the $Q^2$ dependence of the \sigt\ data
at $-t$=0.08 and 0.2 GeV$^2$, scaled to $W$=2.19 GeV. The data show a clear dependence 
on $Q^2$, which is reasonably-well described by a factor of the form $\frac{C}{1+DQ^2}$.

\begin{figure}
\begin{center}
\includegraphics[width=3.5in]{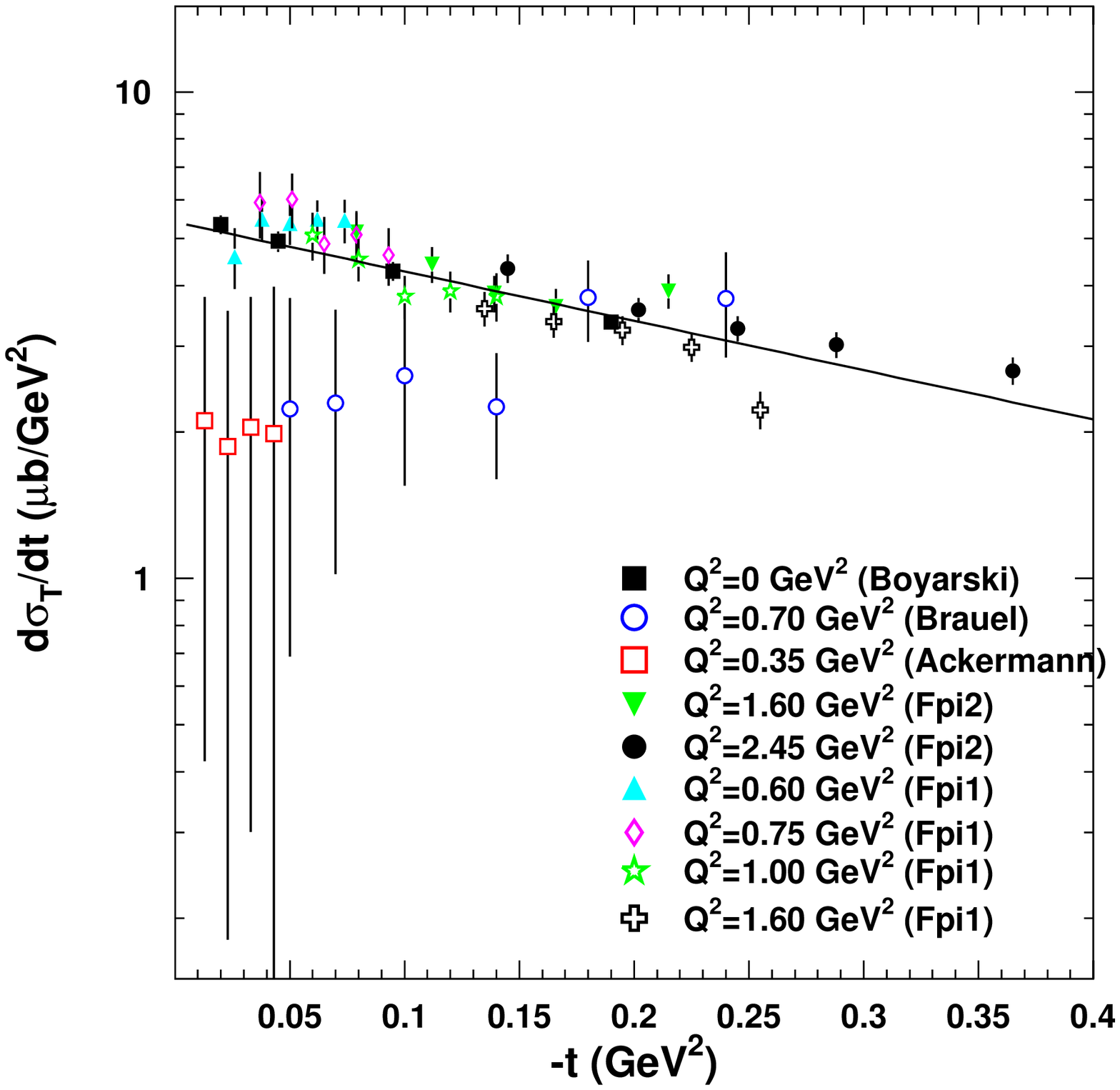}
\caption{\label{fig-sigt-tdep-all} \it (Color online) The $t$ dependence of the transverse
  $\pi^+$ cross section. The data from~\cite{boy68,ack78,bra76,tad06,horn06} 
  are scaled to common values of $W=2.19$ GeV and $Q^2$=0.7 GeV$^2$ (see the text).}
\end{center}
\end{figure}

Fig.~\ref{fig-sigt-tdep-all} displays the electro- and photo-production
\sigt\ data scaled to $W$=2.19 GeV using the functional form $(W^2 - M^2)^{-2}$, 
and to $Q^2$=0.7~GeV$^2$ according to $\frac{C}{1+DQ^2}$, where $C$=8.21$\pm$1.7 and $D$=1.54$\pm$1.7.
An exponential in $t$ analogous to the form used for \sigL\ describes the 
photoproduction, and electroproduction data from Fpi-1 and Fpi-2 to within 30\%, while the DESY data 
are overpredicted by a factor of about two. An exponential fit results in
a slope parameter of $b$=2.3 $\pm$ 1.5 GeV$^{-2}$, and a normalization factor of 
$B$=5.4 $\pm$ 1.4. 
Though the slope is less steep than for \sigL\, it is clear that
\sigT\ is not independent of $t$ and $Q^2$ in these kinematics. This is different
from the conclusions of ~\cite{bra76}.   

\begin{figure}[!htb]
\begin{center}
\includegraphics[width=3.5in]{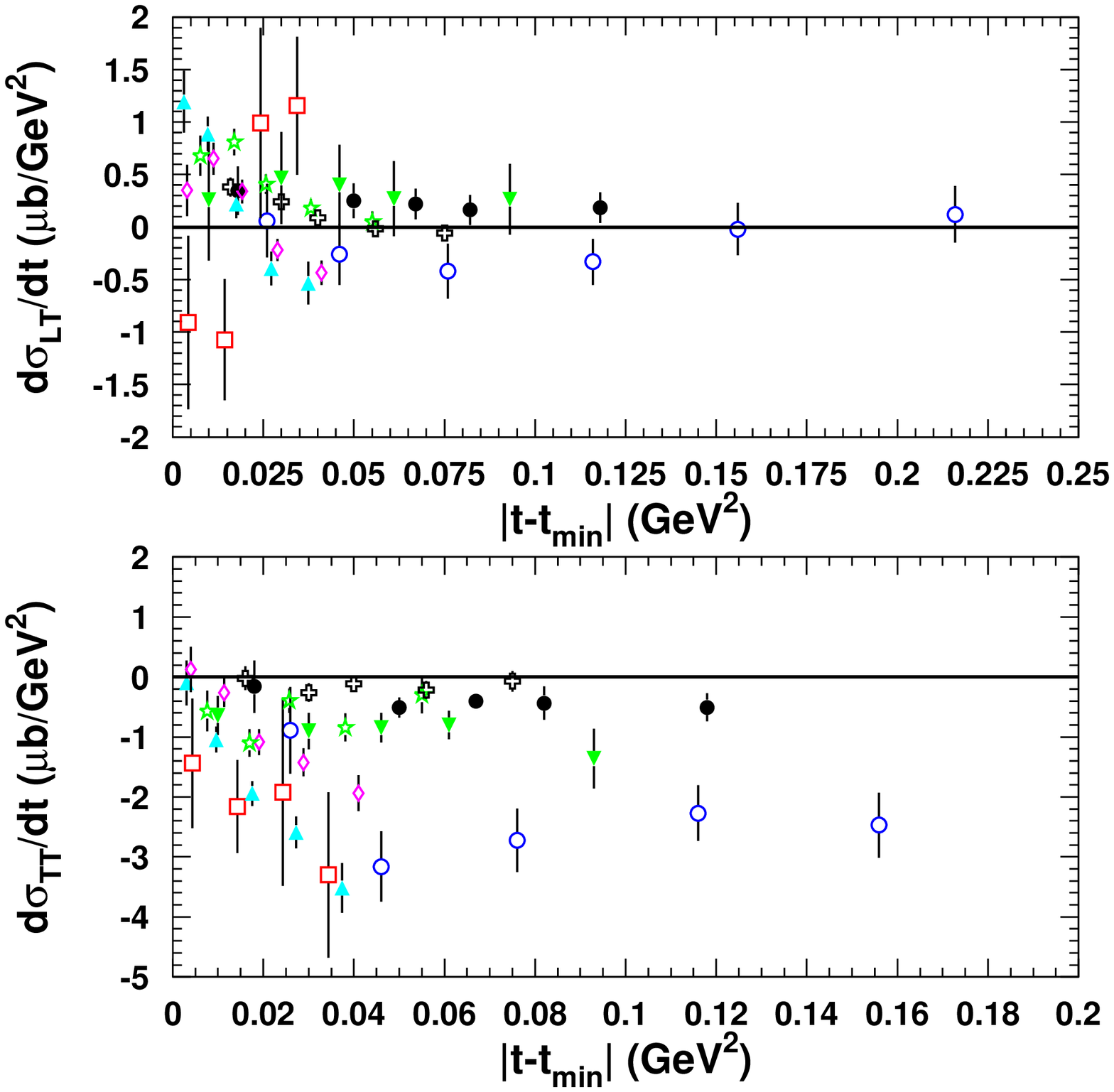}
\caption{\label{fig-siglt-sigtt-tdep-all}
\it (Color online) The $t-t_{min}$ dependence of the interference terms. Both are scaled to common values
of $W$=2.19 GeV and $Q^2$=0.70 GeV$^2$, see the text.
The plotting symbols are the same as in Fig.~\ref{fig-sigt-tdep-all}.}
\end{center}
\end{figure}

Fig.~\ref{fig-siglt-sigtt-tdep-all}
shows the $t-t_{min}$ dependence of \sigLT\ and \sigTT. The \sigLT\ data are scaled to
common values of $W$=2.19 GeV and $Q^2$=0.70 GeV$^2$ using a factor $Q^2 F_{\pi}$.
The \sigTT\ data were scaled in $Q^2$ analogous to \sigT.
In both cases, no overall trend could be identified
due to the large scatter of the data.

\subsection{VGL Regge Model \label{sec:vgl}}

In~\cite{gui97,van97}, Vanderhaeghen, Guidal and Laget (VGL) 
developed a Regge model for pion production, in which the pole-like propagators
of Born term models are replaced with Regge propagators, i.e., the interaction
is effectively described by the exchange of a family of particles with the same
quantum numbers instead of a single particle.  For forward pion
production, the dominant exchanges are the $\pi$ and $\rho$ trajectories. These
determine the $t$-dependence of the cross section without the use of a 
$g_{\pi NN}(t)$ factor.  Since the $t$-channel $\pi$ diagram is by itself not gauge
invariant, in the VGL model the $s$-channel (for $\pi^+$ production) or
$u$-channel (for $\pi^-$ production) nucleon exchange diagram was also
Reggeized, to ensure gauge invariance of their sum.  The model
is parameter free, as the coupling constants at the vertices (such as
$g_{\rho\pi\gamma}$) are well determined by precise studies and analyses in the
resonance region.

The VGL model was first applied to pion photoproduction~\cite{gui97}. 
The model gave a good and consistent description of the
$W$- and $t$-dependences of the available $\pi^+$ and $\pi^-$ photoproduction
data including the spin asymmetries. The fact that both the $\pi$
(unnatural-parity) and the $\rho$ (natural-parity) trajectories are
incorporated in the model proved to be essential to explain the different
behaviors of $\pi^+$ and $\pi^-$ photoproduction.

In~\cite{van97}, the model was extended to pion electroproduction. As the
$\pi$- and $\rho$-exchange amplitudes are separately gauge invariant, two
different electromagnetic form factors were introduced for the $\pi$ and $\rho$
exchanges without violating the gauge invariance of the model. In both cases,
monopole forms are used. Form factors of monopole type were taken for the
$\pi$ and $\rho$ exchanges :
\begin{equation}
\label{eq:monopole}
F_{\pi ,\rho}(Q^2) = [1 + Q^2/\Lambda^2_{\pi ,\rho}]^{-1}.
\end{equation}
The model gave a good description of $\pi^+$ electroproduction data out to
large values of $-t$ at $W$-values of 2.15 and 3.1 GeV for \qsq= 1.2 \gevsq\
~\cite{beb76}, and of the $\pi^-/\pi^+$ ratio at $W=2.19$ GeV, \qsq= 0.7 and
1.35 \gevsq\ ~\cite{bra76}.  

The VGL model is compared to our electroproduction data in Figs.~\ref{fig:xsec1H}
and \ref{fig:xsec2H}.  The VGL cross sections were
evaluated at the same $\overline W$ and $\overline Q^2$ values as the data.
Over the range of $-t$ covered by this work, \sigl\ is completely determined by
the $\pi$~trajectory, while \sigt, \sigtt\ and \siglt\ are also sensitive to
the $\rho$ exchange contribution. Comparison of the model calculations to
previous data gave a value for \Lpi\ of about $0.45-0.50$ \gevsq.  
Here, calculations with a common value of \Lpi=0.462 \gevsq\ are shown.  This
is the same value as is used in~\cite{van97}.
The value of \Lrho\ is more poorly known. 
Here, calculations with \Lrho=1.500 \gevsq\ are shown, where this upper value is determined from the
application of the VGL model to kaon electroproduction~\cite{gui00}.

With a single value of \Lpi=0.462 \gevsq, the VGL model does an overall
good job of describing the magnitude, and $t$, $W$ and $Q^2$-dependences
of our \sigl\ data.  However, as shown in Fig. \ref{fig:xsec1H}, the
description of the $t$-dependence is not as good for $Q^2\leq 1.00$ \gevsq,
$W=1.95$ GeV, where the model prediction is too flat in comparison to the
experimental data.  The model also strongly underestimates \sigt, almost independent
of the value of \Lrho, and this underestimation appears to grow with \qsq, the
fall-off of the data with \qsq\ being less than that of the model.  This
deficiency is also reflected in a too-small prediction for \sigtt. Please note
that VGL's definition of \sigtt\ differs from ours by a minus sign, which has
been included here.  The \siglt\ calculations at $W=2.22$ GeV are generally
satisfactory, but the agreement with the data is much worse at the lower value
$W=1.95$ GeV, the data getting smaller or even becoming negative at larger
values of $-t$.

Recently, the VGL model was extended~\cite{giessen} by including, apart from
a slightly different way to handle the gauge invariance, a hard scattering between
the virtual photon and a quark, followed by hadronization of the system into a pion plus
residual nucleon. With plausible assumptions a good description of \sigt\ was obtained,
with no influence on \sigl.

\subsection{FGLO Effective Lagrangian Model}
\label{sec:fglo}

A more recent development is the effective Lagrangian model of Faessler,
Gutsche, Lyubovitskij and Obukhovsky (FGLO, ~\cite{obu06,obu07}).  This is
a modified Born Term Model, in which an effective Lagrangian is used to describe
nucleon, pion, $\rho$, and photon degrees of freedom.  The (combined) effect of
$s$- and $u$-channel contributions, which interferes with the pion $t$-pole, is
modeled using a constituent quark model.  The authors pay special attention to
the role of the $\rho$ meson in $\pi^+$ electroproduction and show that the
$\rho$ $t$-pole contribution is very important for obtaining a good description
of the magnitude of \sigt.  When comparing vector and tensor representations of
the $\rho$ contribution, the latter was found to give better results.  Unlike
the VGL model, the \sigl\ cross section depends here also on
the $\rho$ exchange, because of the interference of the $\pi$ and tensor $\rho$
exchange contributions.  The model contains a few free parameters, such as the
renormalization constant of the Kroll-Ruderman contact term used to model the
$s(u)$-channel, and $t$-dependent strong meson-nucleon vertices, which are
parameterized in monopole form, as are the electromagnetic form factors.  The
corresponding parameters were adjusted so as to give overall good agreement
with our \sigl\ and \sigt\ data.
\begin{figure}
\begin{center}
\includegraphics[width=3.5in]{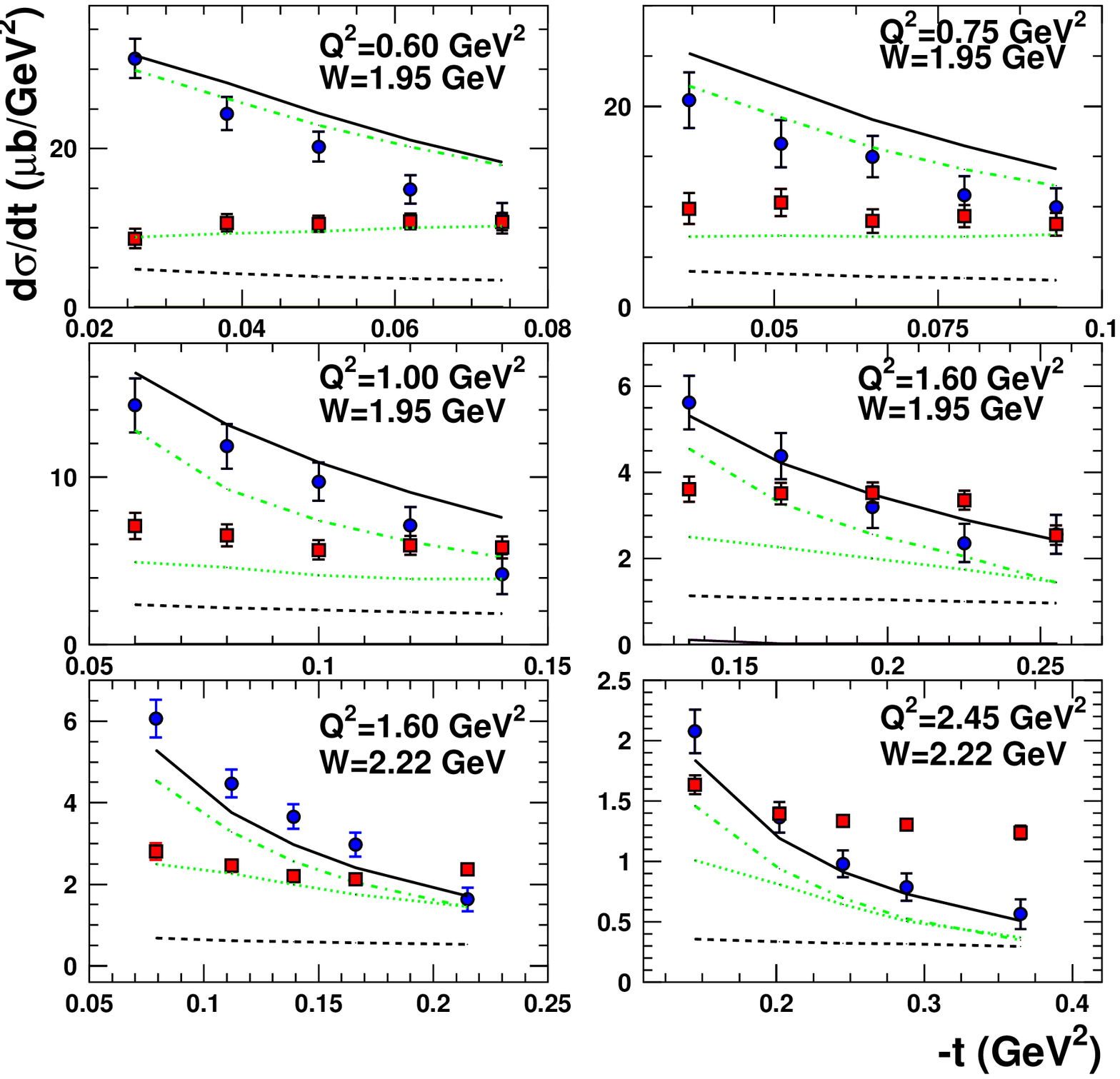}
\caption{\label{fig:xsec_obu} \it Separated $\pi^+$ electroproduction cross
  sections, \sigL\ (circles) and \sigT\ (squares) at central values of $Q^2$=0.60, 0.75,
  1.00, 1.60 GeV$^2$ ($W$=1.95~GeV), and $Q^2$=1.60, 2.45 GeV$^2$ ($W$=2.22~GeV)
  in comparison to the predictions of the VGL Regge~\cite{van97} (solid line) and to the
  FGLO effective Lagrangian~\cite{obu06,obu07} (dashed line) model. A common value of \Lpi=0.462 \gevsq\ is used. 
  Note that the average values
  of $W$ and $Q^2$ are different for each $-t$-bin. The error bars denote
  statistical and $t$ uncorrelated systematic uncertainties combined in
  quadrature. In addition, there is a $t$ and $\epsilon$ correlated systematic
  uncertainty of 4-6\%, by which all data points move collectively.}
\end{center}
\end{figure}

The FGLO model calculation is compared to our data in Fig. \ref{fig:xsec_obu}.
A common value of \Lpi=0.462 \gevsq\ is used throughout.
The other model parameters were fixed at the values assigned by the authors. 
Generally, the agreement of the FGLO model with the \sigl\ data is rather good,
but the model gives a too-flat $t$-dependence at \qsq=0.60 \gevsq, $W=1.95$
GeV.   While on average the model
calculation is in agreement with the \sigt\ data, it fails to describe the
$Q^2$- and $W$-dependences.  For example, the model under-predicts
the \qsq=1.60 \gevsq, $W=1.95$ GeV \sigt\ data by
about a factor of two, while those at \qsq=1.60 \gevsq, $W=2.22$ GeV are
reproduced, and the \qsq=2.45 \gevsq, $W=2.22$ GeV \sigt\ data under-predicted
again by 20-60\%.  
No calculations are available for the
interference cross sections.

\subsection{VGG GPD model}
\label{sec:vgg}

Vanderhaeghen, Guichon, and Guidal (VGG)~\cite{Vdh98} have performed a
calculation for \sigL\ using Generalized Parton Distributions (GPDs). This
approach is based on a soft-hard factorization theorem~\cite{Coll97}. 

Since the 1-gluon perturbative diagram 
severely underestimates the
value of the pion form factor at the relevant $Q^2$, power corrections due to
intrinsic transverse momenta and soft overlap contributions were included in
the calculation, thereby increasing the calculated cross
sections by an order of magnitude. 
The VGG GPD model is compared to our electroproduction data in Fig.~\ref{fig:xsec1H}. 
The GPD calculation gives a
rather good description of the $t$-dependence of the $W>$ 2 GeV data, while the
$Q^2$-dependence is also described fairly well. 
The determination of the onset of this regime remains one of the great 
challenges in contemporary GPD studies. Measurements to address this issue, 
approved for data taking after the completion of the JLab 
upgrade~\cite{E12-07-103}, may be
expected to place a 
constraint on the value of $Q^2$ for which one can reliably apply perturbative 
QCD concepts and extract Generalized Parton Distributions. 

\section{Summary and conclusions}
\label{sec:summary}\nopagebreak[4]

Precision data for the \heepi\ reaction were obtained in order 
to study the pion form factor in the regime \qsq=0.5-3.0 \gevsq.
The data were acquired at JLab making use of the high-intensity, 
continuous CEBAF electron beams and the magnetic spectrometers in Hall C.

The \heepi\ cross sections were measured for values of the Mandelstam variable
$t$ close to its minimum value $t_{min}$ for
(central) four-momentum transfers ranging from $Q^2$=0.60 to 2.45 GeV$^2$, at an 
invariant mass of the photon-nucleon system of $W$=1.95 or 2.22 GeV.
Since \fpi\ is to be determined from the longitudinal part, \sigl, of the
cross section, the measured cross sections were decomposed into
the four structure functions \sigl, \sigt, \siglt, and \sigtt\ at every $Q^2$.
This required measuring the cross section at two values of the virtual photon 
polarization, $\epsilon$, and as function of the azimuthal angle $\phi$ of the 
produced pion.
In the analysis, a Monte Carlo simulation of the whole
experimental setup was used. The simulation included a model cross section fitted to
the data, and thus allowed for accurate acceptance corrections.

Good control of the systematic uncertainty is extremely important in L/T 
separations, as the error bars are inflated by $\Delta \epsilon$.
Therefore, all parts of the experimental setup and the analysis procedures were
carefully inspected and calibrated. This included the optical properties
of the spectrometers, the tracking and particle identification methods,
and the various efficiencies.
As a result, the total systematic uncertainty of the unseparated cross sections
could be reduced to a point-to-point uncertainty below 2\%, plus a scale
uncertainty of less than 3.5\%. The final separated cross sections have a total
uncertainty (statistical plus systematic) between 8 and 15\%. 

The longitudinal cross section \sigL\ shows the characteristic fall-off with
$-t$ due to the pion pole,
and largely behaves as function of \qsq\ according to $Q^2 F_{\pi}(Q^2)^2$.
The transverse cross section \sigT\ depends only little on 
$t$, but our results indicate a clear dependence on \qsq, which, including
photoproduction data, can be described as $\frac{1}{1+bQ^2}$.
This is different from what was concluded from earlier electroproduction
results at DESY.
The interference term \sigLT\ is rather small, while 
the value of \sigTT\ drops fast with increasing \qsq. 

The separated cross sections were compared with the results of model
calculations for the \heepi\ reaction, which use 
Regge trajectories,
effective Lagrangians, or Generalized Parton Distributions. They all
provide a fair to good description of the longitudinal cross section.
The description of the transverse cross section is much worse, however.
The Regge model strongly underpredicts \sigt,
while the Lagrangian model yields good agreement at some values of
$W$ and \qsq, but fails when either $W$ or \qsq\ is varied.
Clearly, more theoretical work has to be done to understand the behavior
of \sigt\ (and also of the interference structure functions \siglt\ and
\sigtt) of the \heepi\ reaction as function of $W$, \qsq\ and $t$.

\section{Acknowledgments}
The authors would like to thank Drs. Guidal, Laget, and Vanderhaeghen for
stimulating discussions and for modifying their computer program for our needs.
We would also like to thank Dr. Obukhovsky for supplying the result of their
model calculations and for many informative discussions.  This work was supported 
in part by the U.S. Department of Energy. The Southeastern Universities Research 
Association (SURA) operates the Thomas Jefferson National Accelerator Facility for 
the United States Department of Energy under contract DE-AC05-84150. We acknowledge 
additional research grants from the U.S. National Science Foundation, the Natural 
Sciences and Engineering Research Council of Canada (NSERC), NATO, FOM (Netherlands), 
and KOSEF (South Korea).

\newpage
\hyphenation{Post-Script Sprin-ger}

\end{document}